\documentclass[fleqn,usenatbib]{mnras}

\usepackage{newtxtext,newtxmath}

\usepackage[T1]{fontenc}
\usepackage{ae,aecompl,times}

\usepackage{graphicx}	
\usepackage{amsmath}	
\usepackage{amssymb}	
\usepackage{multirow}
\usepackage[normalem]{ulem}
\usepackage{xcolor}
\usepackage{bm}

\newcommand{\Mpch}{\,h^{-1}{\rm Mpc}}
\newcommand{\hMpc}{\,h{\rm Mpc}^{-1}}
\newcommand{\Msh}{\,h^{-1}M_\odot}
\newcommand{\lla}{\left\langle}
\newcommand{\rra}{\right\rangle}

\title[Large-scale RSD in MG theories]{Large-scale redshift space distortions in modified gravity theories}

\author[C.~Hern\'andez-Aguayo et al.]
{
\parbox{0.9\textwidth}{C\'esar Hern\'andez-Aguayo$^{1}$\thanks{E-mail: 
cesar.hernandez-aguayo@durham.ac.uk (CH-A)},
Jiamin Hou$^{2}$\thanks{jiamin.hou@mpe.mpg.de (JH)},
Baojiu Li$^{1}$,
Carlton M.~Baugh$^{1}$\\
\& Ariel G.~S\'anchez$^{2}$
}
\\
\\
$^{1}$Institute for Computational Cosmology, Department of Physics, Durham University, South Road, Durham, DH1 3LE, UK.\\
$^{2}$Max-Planck-Institut f\"ur Extraterrestische Physik, Postfach 1312, Giessenbachstr., 85748 Garching, Germany.
}

\date{Accepted XXX. Received YYY; in original form ZZZ}

\pubyear{2018}

\begin{document}
\label{firstpage}
\pagerange{\pageref{firstpage}--\pageref{lastpage}}
\maketitle

\begin{abstract}
Measurements of redshift space distortions (RSD) provide a means to test models of gravity on large-scales. We use mock galaxy catalogues constructed from large N-body simulations of standard and modified gravity models to measure galaxy clustering in redshift space. We focus our attention on two of the most representative and popular families of modified gravity models: the Hu \& Sawicki $f(R)$ gravity and the normal branch of the DGP model. The galaxy catalogues are built using a halo occupation distribution (HOD) prescription with the HOD parameters in the modified gravity models tuned to match with the number density and the real-space clustering of {\sc boss-cmass} galaxies. We employ two approaches to model RSD: the first is based on linear perturbation theory and the second models non-linear effects on small-scales by assuming standard gravity and including biasing and RSD effects. We measure the monopole to real-space correlation function ratio, the quadrupole to monopole ratio, clustering wedges and multipoles of the correlation function and use these statistics to find the constraints on the distortion parameter, $\beta$. We find that the linear model fails to reproduce the N-body simulation results and the true value of $\beta$ on scales $s < 40\Mpch$, while the non-linear modelling of RSD recovers the value of $\beta$ on the scales of interest for all models. RSD on large scales ($s\gtrsim20$-$40\Mpch$) have  been found to show significant deviations from the prediction of standard gravity in the DGP models. However, the potential to use RSD to constrain $f(R)$ models is less promising, due to the different screening mechanism in this model.
\end{abstract}




\section{Introduction}
\label{sec:intro}
Cosmology has entered into a golden era of high precision measurements. New observations will allow us to better understand the nature of the $96\%$ of the energy content of the Universe that is thought to be dark \citep{Hinshaw:2012aka,Planck:cosmoparams}. Current and future galaxy surveys such as the Baryon Oscillation Spectroscopic Survey ({\sc boss};  \citealt{Alam:2017hwk}), the Dark Energy Spectroscopic Instrument ({\sc desi}; \citealt{Flaugher:2014lfa}), the 4-metre Multi-Object Spectroscopic Telescope ({\sc 4most}; \citealt{4most}) and {\sc Euclid} \citep{Laureijs:2011gra,Amendola:2012ys}, aim to measure the spatial distribution of millions and billions of galaxies to reveal the nature of the accelerated expansion of the Universe. 

One complication when using such surveys is that the distance to the galaxies is inferred from their redshifts by assuming a cosmological model, to give positions in ``redshift space''. The peculiar velocities of galaxies along the line of sight (LOS), gravitationally induced motions in addition to the Hubble flow, cause a displacement to the position of the galaxy in redshift space compared to its true position, which is known as the redshift space distortion (RSD) of galaxy clustering. This phenomenon is demonstrated in simulations with ``emulated'' galaxies (see e.g. \citealt{Tinker:2005na,Tinker:2006dm,Kwan:2011hr,Marulli:2015jga}).  
The RSD effect can be combined with other observables such as the baryon acoustic oscillation (BAO) pattern to put constraints on the growth rate of the large-scale structures as well as the cosmological parameters. A wide range of tracers, including luminous red galaxies \citep{Cabre:2008sz,Sanchez:2009}, cosmic voids \citep{Hamaus:2015yza,Hamaus:2017dwj,Cai:2016jek} and quasi-stellar-objects (QSOs) \citep{Hou:2018dr14, Gil-marin:2018dr14, Zarrouk:2018dr14} have been successfully used to extract cosmological information by assuming a standard cosmological model ($\Lambda$CDM) based on General Relativity (GR).

In spite of the success of $\Lambda$CDM, this model does not offer a universally accepted solution to the cosmic acceleration problem. Alternative theories of gravity (commonly referred to as modified gravity theories; MG, for reviews see, \citealt{Copeland:2006wr,Joyceetal2015,Koyama2016}) can have a similar cosmic expansion history to that in $\Lambda$CDM but with different evolution of the growth rate, usually parametrised as $f(z) \simeq \Omega^{\gamma}_{\rm m}(z)$, which depends on the mass density parameter $\Omega_{\rm m}(z)$ and a fitting index $\gamma$ \citep{Linder:2007APh}. A deviation of the index from $\gamma = 0.55$ would indicate a different theory than GR with distinctive gravitational evolution, and therefore has a direct impact on the anisotropic clustering caused by the RSD effect.
Due to the degeneracy between the growth rate and the matter fluctuation amplitude, instead of probing $f(z)$ directly, we use the linear distortion parameter, $\beta(z) = f(z)/b(z)$, where $b(z)$ is the linear tracer bias \citep{Kaiser:1987,Hamilton:1992zz}. Two representative families of MG models are the Hu-Sawicki $f(R)$ gravity model \citep{Hu:2007nk} and the normal branch of the Dvali$-$Gabadadze$-$Porrati model (nDGP; \citealt{Dvali:2000hr}) which, as we shall see later, make distinct predictions for the linear growth rate $f(z)$. 

The impact of modified gravity on RSD has been studied in a number of previous works. For instance, \citet{Jennings:2012pt} and \citet{Xu:2014wda} presented  predictions for the RSD in $f(R)$ gravity in Fourier space and \citet{Arnalte-Mur:2016alq} complemented these results in configuration space by measuring the correlation function. These studies are based either on the matter or halo density fields as predicted using N-body simulations of modified gravity, or using analytical fitting formulae for matter clustering which are themselves derived from simulations. This demonstrates the importance of using cosmological simulations to study the RSD effect on mildly and strongly nonlinear scales. Recently, \citet{He:2018oai} used high-resolution simulations of $\Lambda$CDM and $f(R)$ gravity to study the small-scale RSD effect. These authors found that the predictions of $f(R)$ gravity are strongly disfavoured by current measurements of galaxy clustering in redshift-space on scales between $1\sim10\Mpch$, while the data is in excellent agreement with $\Lambda$CDM. Arguably, simulations are the only way to accurately predict RSD effects down to such small scales, but the main disadvantage of this approach is the high cost of running large suites of high-resolution simulations to explore the parameter space, and the uncertainties in the connection between galaxies and simulated dark matter haloes. Regarding the galaxy - dark matter halo connection, hydrodynamical simulations and semi-analytical modelling, two approaches which add elements of the physics of galaxy formation to the modelling of the growth of structure in the dark matter,  will inform empirical treatments of the galaxy-halo connection, such as the one used in this paper (e.g. \citealt{Contreras:2013kr,Chaves-Montero:2015iga,Desmond:2017ghc}).

In the mean time, the importance of the theoretical modelling of RSD in modified gravity is being realised and increasing effort is being devoted to improving the description of the RSD effects in $f(R)$ gravity and nDGP models on mildly nonlinear scales, based on higher-order perturbation theory \citep[see, e.g., the pioneering works of][]{Taruya:2013quf,Bose:2016qun,Bose:2017myh,Bose:2017dtl}. In particular, the theoretical modelling presented by \citet{Bose:2016qun} and \citet{Bose:2017dtl} has been carefully compared against N-body simulations and found to show good agreement with the results for the power spectrum and correlation function in real and redshift space. These authors have gone to great lengths to include higher-order perturbation terms in the description of the MG effects  to ensure that the latter are modelled consistently and accurately. These analytical or semi-numerical approaches are much more efficient than full N-body or hydrodynamical simulations, though their validity is usually restricted to mildly nonlinear scales, such as $k\leq0.1-0.3\hMpc$.

Here, we explore the following questions: (1) Do galaxy catalogues from the current and next generations of galaxy surveys offer the realistic possibility to constrain or rule out some of the leading modified gravity models in the literature using RSD? (2) Given reasonable estimates of theoretical and observational uncertainties, is a simpler treatment of the RSD effect (which ideally does not involve new theoretical templates -- based on simulations or perturbation theory --  to be developed each time the gravity model or its parameter is changed)  sufficient to measure $\beta$ and distinguish between models? The simpler treatments we consider include (i) a linear theory model \citep{Kaiser:1987,Hamilton:1992zz} and (ii) a model that accounts for non-linear matter evolution following the approach of Crocce, Blas \& Scoccimarro (in prep.), who extend renormalised perturbation theory \citep[RPT,][]{Crocce:2005xy} by using Galilean invariance to find a resummation of the mode-coupling power spectrum, galaxy bias as in \citet{Chan:2012jj}, and a detailed description of RSD \citep{Scoccimarro:2004tg}, which does {\it not} incorporate any MG effect. The approach we take is to directly confront the RSD predicted by these simplified treatments against RSD measurements from mock galaxy catalogues built on simulations of different gravity models, and check if the input $\beta$ values can be faithfully recovered. The linear theory prediction is model-independent, and the nonlinear model used is for GR only so that it is also effectively model-independent

Recently, \citet{Barreira:2016ovx} used the same model for non-linearities, bias, and RSD to estimate the growth rate in nDGP models, also using mock galaxy catalogues built from N-body simulations. There are several differences in the work presented here compared with that of \cite{Barreira:2016ovx}. Firstly, we consider a wider variety of models by including also variants of $f(R)$ gravity and different parameter values for nDGP. Secondly, we conduct an explicit comparison of linear Kaiser and nonlinear models, considering different estimators and ranges of scales in the parameter fitting to test for systematic effects. Thirdly, the mock galaxy catalogues used here are constructed in a different way from that used by \citet{Barreira:2016ovx}, using larger simulations.

The paper is organised as follows. In Sect.~\ref{sec:MG} we introduce the theory of the modified gravity models considered. Section \ref{sec:simulation} describes the N-body simulations and the construction of mock galaxy catalogues in real and redshift space. In Sect.~\ref{sec:RSD} we outline the theoretical aspects of redshift space distortions. Results from the linear and non-linear RSD models are presented in \ref{sec:Kaiser} and \ref{sec:gRPT} of Section \ref{sect:model_constraints}, respectively, and we discuss the results in Section \ref{sec:result_discuss}. Finally, Sect.~\ref{sec:conc} contains the summary and conclusions. 

Throughout this paper we use the convention that the speed of light $c=1$, and Greek indices $\mu,\nu,\cdots,$ run over $0,1,2,3$.

\section{Modified gravity theory}\label{sec:MG}
We start with a brief introduction to the $f(R)$ and nDGP models of modified gravity (Sec.~\ref{sec:models}), then give the equations in these models that govern non-linear structure formation (Sec.~\ref{sec:struc}) and finally briefly explain the screening mechanisms necessary to suppress the effects of modified gravity and recover GR in regions such as the Solar System (Sec.~\ref{sec:screen}).

\subsection{Models}\label{sec:models}
\begin{figure*}
 \centering
\includegraphics[width=0.45\textwidth]{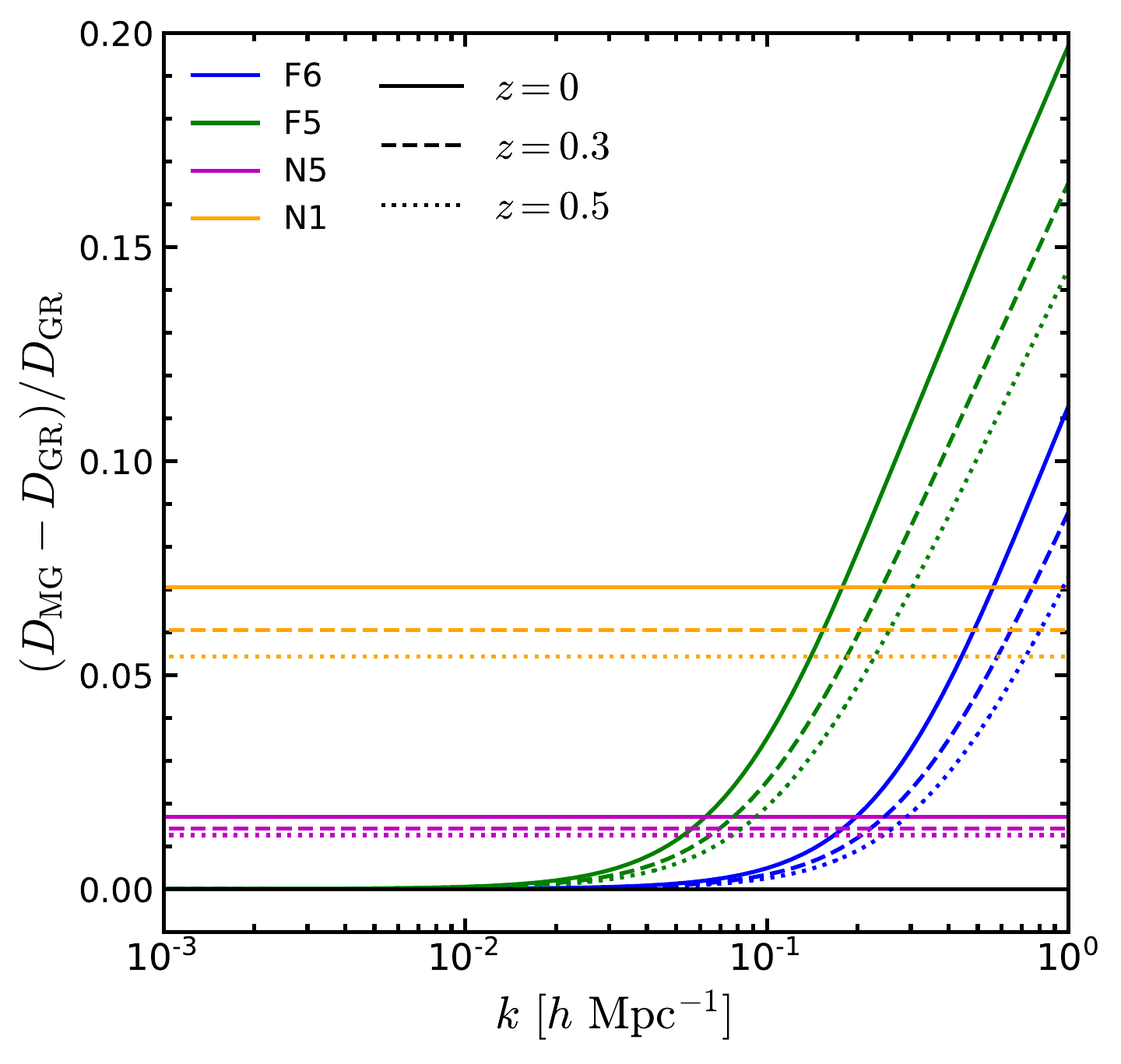}
\includegraphics[width=0.45\textwidth]{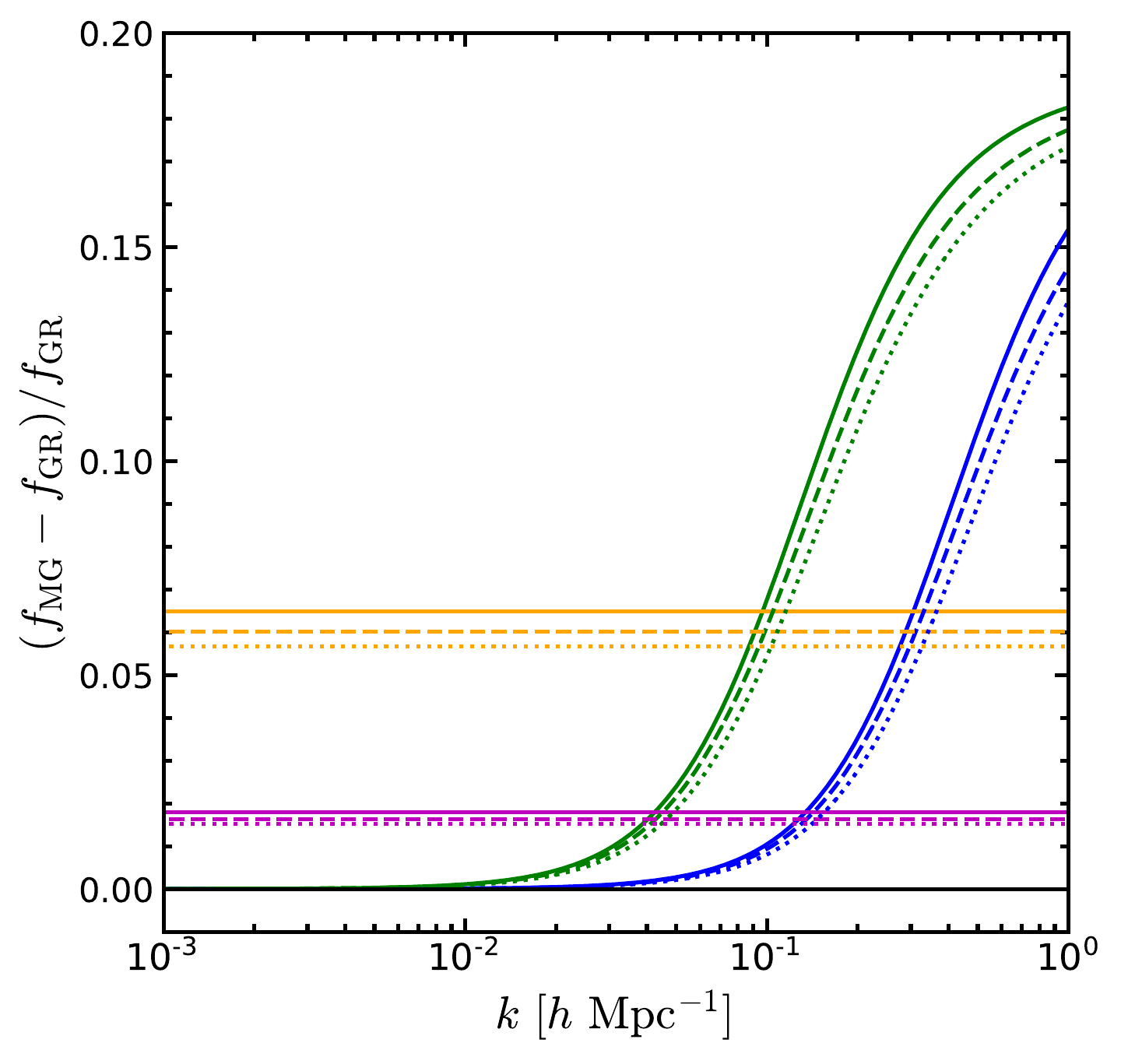}
	\caption{Relative difference of the linear growth factor ($D$ Eq.~\eqref{eq:Dp}, {\it left panel}) and the linear growth rate ($f$ Eq.~\eqref{eq:f_lin}, {\it right panel}) between different gravity models (F6, F5, N5 and N1) and GR at different redshifts as a function of the wavenumber, $k$. The colour scheme and line styles are specified in the legend and show different models and redshifts.}
	\label{fig:f_D_z}
\end{figure*}
\subsubsection{$f(R)$ gravity}\label{sec:fR}

The modified Einstein equations in $f(R)$ gravity can be obtained by varying the modified Einstein-Hilbert action
\begin{equation}\label{eq:S-f(R)}
S = \frac{1}{16\pi G} \int \mathrm{d}^4x\sqrt{-g} (R + f(R)) + S_{\rm m}(g_{\mu\nu},\psi_i)\, ,
\end{equation}
with respect to the metric, $g_{\mu\nu}$, 
\begin{equation}\label{eq:M-EQ}
G_{\mu\nu} + f_R R_{\mu\nu} - g_{\mu\nu} \left(\frac{1}{2}f(R) - \Box f_R \right) - \nabla_\mu \nabla_\nu f_R = 8\pi G T^{\rm m}_{\mu\nu}\,,
\end{equation}
where $R$, $R_{\mu\nu}$ and $G_{\mu\nu}$ are respectively the Ricci scalar, Ricci tensor and Einstein tensor, $\nabla_\mu$ is the covariant derivative, $\Box = \nabla^\mu \nabla_\mu$ the d'Alambertian, $G$ is the gravitational constant, $g$ is the determinant of the metric, $S_{\rm m}$ is the action of the matter fields $\psi_i$ and $T^{\rm m}_{\mu \nu}$ is the energy-momentum tensor for matter.

Eq.~\eqref{eq:M-EQ} contains a new dynamical degree of freedom, known as the \emph{scalaron} field and defined by
\begin{equation}\label{eq:sc}
f_R \equiv \frac{\mathrm{d}f(R)}{\mathrm{d}R}\,. 
\end{equation}
The amplitude of this scalaron field determines the deviations from GR, with larger $|f_R|$ meaning stronger deviations.

The evolution of the scalaron field is obtained by taking the trace of the modified Einstein equations, Eq.~\eqref{eq:M-EQ}, 
\begin{equation}\label{eq:f_R}
\Box f_R = \frac{1}{3} (R - f_R R + 2f(R) + 8\pi G\rho_{\rm m})\,,
\end{equation}
where $\rho_{\rm m}$ is the non-relativistic matter density of the Universe.

Various functional forms of $f(R)$ have been proposed in the literature to study modifications to general relativity (for reviews see e.g. \citealt{DeFelice:2010aj,Sotiriou:2010rp}). Here we consider the \citet*{Hu:2007nk} model
\begin{equation}\label{eq:f(R)}
f(R) = -m^2 \frac{c_1}{c_2} \frac{(-R/m^2)^n}{(-R/m^2)^n + 1}\,,
\end{equation}
\begin{equation}\label{eq:sc1}
f_R = -\frac{c_1}{c^2_2} \frac{n(-R/m^2)^{n-1}}{[(-R/m^2)^n + 1]^2}\,,
\end{equation}
where $m$ is a new mass scale defined as $m^2 \equiv H^2_0 \Omega_{\rm m}$, $H_0$ is the current value of the Hubble expansion rate, $\Omega_{\rm m}$ is the current density parameter of non-relativistic matter, $n$, $c_1$ and $c_2$ are free dimensionless parameters of the model. 
This specific $f(R)$ model can approximately mimic the background expansion of the $\Lambda$CDM model if we fix $c_1/c_2 = 6(\Omega_\Lambda/\Omega_{\rm m})$, where $\Omega_\Lambda \equiv 1 - \Omega_{\rm m}$. 

Eqn.~\eqref{eq:sc1} can be approximated as
\begin{equation}\label{eq:sc2}
f_R \approx -n\frac{c_1}{c^2_2} \left(\frac{m^2}{-R} \right)^{n+1}\,,
\end{equation}
in the limit $|\bar{R}| \approx 40 m^2 \gg m^2$, a condition that is satisfied throughout the cosmic history with reasonable parameter values $\Omega_{\rm m} \approx 0.3$ and $\Omega_\Lambda \approx 0.7$, with
\begin{equation}\label{eq:R}
-\bar{R} \approx 8\pi G\bar{\rho}_{\rm m} - 2f(\bar{R}) \approx 3 m^2 \left[ a^{-3}  + \frac{2}{3} \frac{c_1}{c_2}\right]\,,
\end{equation}
where $a$ is the scale factor, normalised to $a = 1$ at the present time.

From the functional form of the scalaron field, Eq.~\eqref{eq:sc1}, we can see that this model has two free parameters, $n$ and $c_1/c^2_2$. In the literature it is common to use $f_{R0}$, which has the physical meaning of being the value of $f_R$ today, instead of $c_1/c^2_2$, where 
\begin{equation}\label{eq:c12}
\frac{c_1}{c^2_2} = -\frac{1}{n} \left[ 3\left(1 + 4\frac{\Omega_\Lambda}{\Omega_{\rm m}} \right) \right]^{n+1} f_{R0}\,.
\end{equation}
Therefore, a particular choice of $n$ and $f_{R0}$ fully specifies the Hu-Sawicki $f(R)$ model. In this work we focus on the cases of $n=1$ and $f_{R0} = -10^{-6}, -10^{-5}$, referred as F6 and F5, respectively. 

\subsubsection{Dvali-Gabadadze-Porrati model}\label{sec:nDGP}

In the \citeauthor*{Dvali:2000hr} (DGP) braneworld model, the Universe is a four-dimensional brane that is embedded in a five-dimensional spacetime (the bulk). The gravitational action of the model is given by
\begin{eqnarray}
S = \int_{\rm brane} {\rm d}^4x \sqrt{-g} \left(\frac{R}{16\pi G}\right) + \int_{\rm bulk} {\rm d}^5x \sqrt{-g^{(5)}} \left(\frac{R^{(5)}}{16\pi G^{(5)}}\right), \label{eq:S_dgp}
\end{eqnarray}
where $g$, $R$ and $G$ have the same meaning as before on the 4D brane, while $g^{(5)}$, $R^{(5)}$ and $G^{(5)}$ are respectively their equivalents in the 5D bulk. A new parameter can be defined as the ratio of $G^{(5)}$ and $G$ and is known as the crossover scale, $r_c$,
\begin{equation}\label{eq:rc}
r_c \equiv \frac{1}{2} \frac{G^{(5)}}{G}.
\end{equation}

In this work we focus on the normal branch DGP (nDGP) model \citep{Sahni:2002dx,Lombriser:2009xg,Schmidt:2009sv}, where the variation of the action, Eq.~\eqref{eq:S_dgp}, yields the modified Friedmann equation 
\begin{eqnarray}\label{eq:H_ndgp}
\frac{H(a)}{H_0} = \sqrt{\Omega_{\rm m}a^{-3} + \Omega_{\Lambda}(a) + \Omega_{\rm rc}} - \sqrt{\Omega_{\rm rc}}\,,
\end{eqnarray}
in a homogeneous and isotropic universe with $\Omega_{\rm rc} = 1/(4H^2_0r^2_c)$. In this model, departures from GR can be quantified by the parameter $H_0r_c$. As we can see from Eq.~\eqref{eq:H_ndgp} if $H_0r_c \rightarrow \infty$ then the expansion of the Universe is closer to $\Lambda$CDM. Therefore, here we work with two nDGP models with $H_0r_c = 5$ and $H_0r_c = 1$ which hereinafter are referred as to N5 and N1 which represent a weak and medium deviation from GR, respectively. 

\subsection{Structure formation}\label{sec:struc}

Since we are interested in the growth of structure in different gravity models, we work with the perturbed Friedmann-Robertson-Walker (FRW) metric in the Newtonian gauge
\begin{equation}\label{eq:FRW}
\mathrm{d}s^2 = (1+2\Psi) \mathrm{d}t^2 - a^2(t)(1-2\Phi)\gamma_{ij}\mathrm{d}\vec{x}^2\,,
\end{equation}
where $\Psi$ and $\Phi$ are the gravitational potentials, $t$ is the cosmic time, and $\vec{x}$ represents comoving coordinates.

In the case of $f(R)$ gravity, non-linear structure formation is determined by the following equations in the quasi-static and weak-field approximations which are known to be good approximations for the regime we are interested in \citep{Bose:2014zba}
\begin{equation}\label{eq:Phi}
\nabla^2 \Phi = \frac{16\pi G}{3} a^2 (\rho_{\rm m} - \bar{\rho}_{\rm m}) + \frac{1}{6} a^2 (R(f_R) - \bar{R})\,,
\end{equation}
for $\Phi$ and
\begin{equation}\label{eq:fR}
\nabla^2 f_R = -\frac{a^2}{3} [R(f_R) - \bar{R} + 8\pi G(\rho_{\rm m} - \bar{\rho}_{\rm m})],
\end{equation}
for $f_R$.
 
On the other hand, structure formation in the nDGP model is governed by the following equations in the quasi-static and weak-field limits \citep*{Koyama:2007ih}
\begin{equation}
\nabla^2 \Phi = 4\pi G a^2 \delta \rho_{\rm m} + \frac{1}{2}\nabla^2\varphi\,,
\end{equation}

\begin{equation}\label{eq:phi_dgp}
\nabla^2 \varphi + \frac{r_c^2}{3\beta_{\rm DGP}\,a^2} \left[ (\nabla^2\varphi)^2
- (\nabla_i\nabla_j\varphi)^2 \right] = \frac{8\pi\,G\,a^2}{3\beta_{\rm DGP}} \delta\rho_{\rm m}\,,
\end{equation}
where $\varphi$ is a scalar degree of freedom related to the bending modes of the brane, and we have defined $\delta\rho_{\rm m} = \rho_{\rm m} - \bar{\rho}_{\rm m}$ as the perturbation of non-relativistic matter. $\beta_{\rm DGP}=\beta_{\rm DGP}(a)$ is a time-dependent function depending on the parameter $r_c$ and the background expansion history:
\begin{equation}\label{eq:beta_dgp}
\beta_{\rm DGP}(a) = 1 + 2 H\, r_c \left ( 1 + \frac{\dot H}{3 H^2} \right ).
\end{equation}
Note that we have used a subscript to distinct this from the $\beta$ parameter introduced in Sect.~\ref{sec:intro}.

The linear growth for the matter fluctuations in these gravity models can be obtained by solving the equation of the linear growth factor, $D$,
\begin{equation}\label{eq:Dp}
D'' + \left(2 - \frac{3}{2}\Omega_{\rm m}(a) \right)D' - \frac{3}{2}\frac{G_{\rm eff}}{G}\Omega_{\rm m}(a) D = 0\,,
\end{equation}
where $^\prime$ denotes a derivative with respect $\ln a$ and $G_{\rm eff}$ takes values of
\begin{equation}\label{eq:Geff}
\frac{G_{\rm eff}}{G} = 
\left \{
      \begin{array}{lc}
          1 & {\rm GR}\,,\\
          1 + k^2/[3(k^2 + a^2m^2_{f_R})] & f(R)\,,\\ 
          1 + 1/[3\beta_{\rm DGP}(a)] & {\rm nDGP} \,,
      \end{array}
   \right.
\end{equation}
here $k$ is the wavenumber, $m_{f_R}$ is the mass of the scalaron field defined by $m^2_{f_R} \simeq [3f_{RR}]^{-1}$. 
Note that $G^{f(R)}_{\rm eff}$ is a function of time and scale, which means that the linear growth of structure for $f(R)$ gravity is scale dependent, while for GR and nDGP is scale independent. 

In galaxy surveys we can extract information about the growth of structure through the linear growth rate, $f$, which is defined as 
\begin{equation}\label{eq:f_lin}
f \equiv \frac{{\rm d}\ln D}{{\rm d}\ln a}.
\end{equation}

Fig.~\ref{fig:f_D_z} shows the relative difference of the linear growth factor, $D$, and the linear growth rate, $f$, at the three redshifts of interest, $z=0,\,0.3,\,0.5$, between different modified gravity models and GR ($\Lambda$CDM) as a function of scale, $k$. The relative differences for the nDGP models remain constant because $D_{\rm nDGP}$ and $f_{\rm nDGP}$ are scale-independent. In the case of $f(R)$ gravity, the difference with respect to GR becomes larger at smaller scales $(k > 0.1 \hMpc)$ and lower redshifs, while at $k < 0.01 \hMpc$ the growth of structure is almost indistinguishable from that in GR.

\subsection{Screening mechanisms}\label{sec:screen}
Modifications to general relativity can lead to interesting effects on all scales. In order to satisfy Solar System constraints these modifications should be hidden in the local environment, hence screening mechanisms have been proposed to recover GR predictions in high-density regions.

In $f(R)$ gravity, the chameleon mechanism \citep{Khoury:2003rn} is introduced to suppress the enhancement of gravity under certain environmental conditions. Since the scalaron field is massive, with a mass given by
\begin{equation}\label{eq:m_eff}
m^2_{f_R} \equiv \frac{\mathrm{d}}{\mathrm{d} f_R}\left( \frac{\mathrm{d}V_\mathrm{eff}}{\mathrm{d} f_R}\right) \simeq \frac{1}{3f_{RR}},
\end{equation}
where the effective potential $V_{\rm eff}$ is defined such that ${\mathrm{d}V_\mathrm{eff}}/{\mathrm{d} f_R} = \Box f_R$; the second equality comes from applying this definition to Eq.~(\ref{eq:f_R}). Hence, for $f_{RR}>0$, the effective potential $V_{\rm eff}$ has a minimum at $f_{R,{\rm min}}$ satisfying $\partial V_{\rm eff}(f_{R,{\rm min}})/\partial f_R=0$. In high-density regions, where $\rho_{\rm m}$ is large, it can be shown, using the expressions of $f(R), f_R$ given above, that $m_{f_R}$ becomes heavy in such that the fifth force decays exponentially as $r^{-2}\exp(-m_{f_R} r)$, leading to recovery of GR. In low-density regions, the fifth force can propagate a further distance, modifying the force law between matter particles. This environmental dependence of the fifth force behaviour earns the screening mechanism the name `chameleon'. 

The nDGP model employs the Vainshtein screening mechanism \citep{Vainshtein:1972sx}. This screening mechanism can be understood by considering spherical symmetry and integrating Eq.~\eqref{eq:phi_dgp} we get
\begin{equation}\label{eq:phi_SS}
\frac{2r_c}{3\beta_{\rm DGP}} \left(\frac{\varphi,_{r}}{r}\right)^2 + \left(\frac{\varphi,_{r}}{r}\right) = \frac{2}{3\beta_{\rm DGP}}\frac{GM(r)}{r^3}\,,
\end{equation}
where $M(r)$ is the mass enclosed within a radius $r$. The solution of Eq.~\eqref{eq:phi_SS} is
\begin{equation}\label{eq:phi_r}
\varphi,_r = \frac{4}{3\beta_{\rm DGP}}\left(\frac{r}{r_V}\right)^3\left[-1 + \sqrt{1 + \left(\frac{r_V}{r}\right)^3}\right]\frac{GM(r)}{r^2}\,,
\end{equation}
where
\begin{equation}\label{eq:r_V}
r_V(r) = \left(\frac{16r^2_c GM(r)}{9\beta_{\rm DGP}^2}\right)^{1/3}\,,
\end{equation}
is the so-called Vainshtein radius. From Eq.~\eqref{eq:r_V}, we can see that $r_V$ offers a scale below which the effects of the fifth force are suppressed. For example, if we consider the range $r \gg r_V > R_{\rm th}$ we find
\begin{equation}
\frac{\varphi,_r}{\Phi_{\rm N},_r} = \frac{2}{3\beta_{\rm DGP}}\,,
\end{equation}
where $\Phi_{\rm N},_r = GM_{\rm th}/r^2$, $R_{\rm th}$ is the radius of a top-hat density profile with mass $M_{\rm th}$. Hence, the fifth force has a strength comparable to the standard Newtonian force (assuming $\beta_{\rm DGP}\sim\mathcal{O}(1)$). Considering the opposite limit when $R_{\rm th} < r \ll r_V $  we have
\begin{equation}
\frac{\varphi,_r}{\Phi_N,_r} \rightarrow 0\, \qquad {\rm as} \qquad \frac{r}{r_V} \rightarrow 0.
\end{equation}
Therefore, the fifth force is suppressed allowing the model to recover the behaviour of GR near massive objects (within their Vainshtein radii) and to pass Solar System tests.

\section{N-body simulations and galaxy catalogues}\label{sec:simulation}
N-body cosmological simulations have played an important role in the analysis of alternative gravity models to GR. Nowadays, these simulations are necessary for the construction of synthetic galaxy catalogues and study the impact of modifications of gravity on the distribution and clustering of galaxies. In this section we present the technical details of the simulations we use and the prescription we follow to generate mock catalogues.

\subsection{N-body simulations of modified gravity}\label{sec:sims_MG}
We use the Extended LEnsing PHysics using ANalaytic ray Tracing ({\sc elephant}) dark matter only N-body simulations which have been run using the {\sc ecosmog} \citep{Li:2011vk} and {\sc ecosmog-v} \citep{Li:2013nua} codes for $f(R)$ gravity and nDGP models respectively. {\sc ecosmog} and {\sc ecosmog-v} are modified versions of the publicly available N-body and hydrodynamical simulation code {\sc ramses} \citep{Teyssier:2001cp}. These codes are efficiently optimised and implemented with methods that speed up the calculations of the non-linear partial differential equations that characterize these models \citep{Barreira:2015xvp,Bose:2016wms}.

The cosmological parameters are the best-fit values from the {\sc wmap9} collaboration \citep{Hinshaw:2012aka} $$\{\Omega_{\rm b}, \Omega_{\rm CDM}, h, n_s, \sigma_8\} = \{0.046,0.235,0.697,0.971,0.82\}.$$ The simulations follow the evolution of $N_{\rm p} = 1024^3$ particles with mass $m_{\rm p}=7.798\times 10^{10}\Msh$ in a cubical box of comoving size $L_\mathrm{box} = 1024\Mpch$ from their initial conditions (generated with the {\sc MPgrafic} code, \citealt{Prunet:2008fv}) at $z_{\rm ini}=49$ up to today $(z_{\rm fi} = 0)$. In this work we used five independent realisations of the matter field for each gravity model. For each realisation the simulations of all gravity models start from the same initial condition, because at $z_{\rm ini}=49$ the effects of modified gravity on large-scale structure formation were negligible for all MG models considered here.

Halo catalogues for all gravity models were constructed using the \textsc{rockstar} halo finder \citep{Behroozi:2011ju} at $z=0,\,0.3$ and $0.5$. We chose $M_{200c}$ as the halo mass definition, which is the mass enclosed within a sphere of radius $r_{200c}$ with $200$ times the critical density of the Universe.

The {\sc elephant} simulations have been used to study the properties of voids for chameleon and Vainshtein mechanism models \citep{Cautun:2017tkc,Paillas:2018wxs} and the halo and galaxy marked correlation functions \citep{Joaquin:2018,Hernandez-Aguayo:2018yrp} in $f(R)$ gravity.

\subsection{Mock galaxy catalogues}\label{sec:HOD}

The next step to measure the impact of modified gravity on redshift space distortions is the generation of mock galaxy catalogues. For this purpose, we built the catalogues by implementing a five-parameter halo occupation distribution (HOD) \citep{Zheng:2004id,Zheng:2007zg} model. This HOD model determines the numbers of central $(\lla N_{\rm c} \rra)$ and satellite $(\lla N_{\rm s} \rra)$ galaxies inside dark matter haloes as functions of the halo mass $(M=M_{200c})$ by following a distribution given by,  
\begin{eqnarray}\label{eq:Ng}
\lla N_{\rm c}(M) \rra &=& \frac{1}{2} \left[ 1 + \mathrm{erf}\left( \frac{\log_{10} M - \log_{10} M_\mathrm{min}}{\sigma_{\log M}} \right) \right]\,,\label{eq:Nc}\\
\lla N_{\rm s}(M) \rra &=& \lla N_{\rm c}(M) \rra \left( \frac{M - M_0}{M_1} \right)^{\alpha}\,.\label{eq:Ns}
\end{eqnarray}
The mean total number of galaxies in each halo is given by $\lla N_{\rm t} \rra = \lla N_{\rm c} \rra + \lla N_{\rm s} \rra$. As we can see from Eq.~\eqref{eq:Nc}, $M_{\rm min}$ and $\sigma_{\log M}$ determine the occupancy of central galaxies while the whole set of parameters determine the mean number of satellite galaxies in each halo (see Eq.~\eqref{eq:Ns}). 

We follow the same prescription as \citet{Manera:2012sc} and \citet{Hernandez-Aguayo:2018yrp} to construct mock catalogues in real space. In summary, when a central galaxy is placed inside a halo we assume that this galaxy is located at the centre of mass of its host halo and takes its coordinates and velocity information. Satellite galaxies (which orbit around central galaxies in haloes with $M_{200c} \ge M_0$) are radially distributed following an Navarro-Frenk-White (NFW) profile \citep{Navarro:1995iw,Navarro:1996gj}, with a uniform angular distribution. The position of satellite galaxies is randomly chosen within the halo radius $(0 < r <r_{200c})$, and their velocity is the halo velocity plus a perturbation along the $x$, $y$ and $z$ coordinates drawn from a Gaussian distribution with variance equal to the 1D velocity dispersion of the host halo.

One of the key steps of the HOD approach is to set the HOD parameters in Eqs.~\eqref{eq:Nc} and \eqref{eq:Ns} to reproduce the galaxy clustering in real galaxy surveys. In addition, given that we only observe one Universe, we need to demand that all galaxy catalogues from all gravity models are consistent with observations. For these reasons, in this work the MG HOD parameters have been calibrated to match with the galaxy number density and the real-space two-point correlation function (which is directly related to the projected correlation function) of the same galaxy sample. In practice, in the case of GR, we take the values of the parameters from the {\sc boss cmass dr9} sample \citep{Manera:2012sc}: $\log_{10}(M_{\rm min}/[h^{-1}M_\odot]) = 13.09$, $\log_{10}(M_1/[h^{-1}M_\odot]) = 14.00$, $\log_{10}(M_0/[h^{-1}M_\odot]) = 13.077$, $\sigma_{\log M} = 0.596$ and $\alpha = 1.0127$. Note that the {\sc cmass} sample has a redshift distribution between $0.4<z<0.7$, which is compatible with one of the three redshifts considered in this work ($z=0.5$); however, we adopt the same HOD parameter values for GR at the other two redshifts ($z=0.3$ and $z=0$) as well, as our objective is to study the measurement of growth rate using RSD for galaxy catalogues with similar real-space clustering, rather than to make precise mock galaxy catalogues for the different gravity models (the latter will be left for future studies). 

To calibrate the HOD parameters for the various MG models, we use a simplex algorithm \citep{Nelder:1965}. The algorithm starts with an initial guess of the HOD parameters, then the code walks through the 5D HOD parameter space looking for the values that minimise the root-mean-square difference of the real-space two-point correlation functions, $\Delta_{\rm rms}$, between models. The relative difference of the galaxy number density $(\Delta n = n_{\rm MG}/n_{\rm GR} - 1)$ is added to $\Delta_{\rm rms}$ to ensure that all models have similar number of tracers. We stop the search when $\Delta_{\rm rms} \leq 0.02$-$0.03$. For more details, see \citet{Li:2017xdi}.

Finally, we use the distant-observer approximation to shift the positions of galaxies from real to redshift space. We use the three coordinates, $\hat{x}$, $\hat{y}$ and $\hat{z}$, as the line-of-sight (LOS) to generate three redshift-space catalogues for one real-space catalogue where the new coordinates are,
\begin{equation}\label{eq:coord_z}
{\bf s} = {\bf r} + \frac{(1+z)v_\parallel}{H(z)}\hat{e}_\parallel\,,
\end{equation}
where ${\bf r}$ is the coordinate vector in real space, $z$ is the redshift of the snapshot used to generate the catalogues, $H(z)$ is the Hubble parameter as a function of $z$, $v_\parallel$ is the component of the velocity along the LOS and $\hat{e}_\parallel$ is the unitary vector of the LOS direction. So, in total we have fifteen redshift space catalogues for each gravity model and each redshift.
\begin{figure}
 \centering
\includegraphics[width=0.45\textwidth]{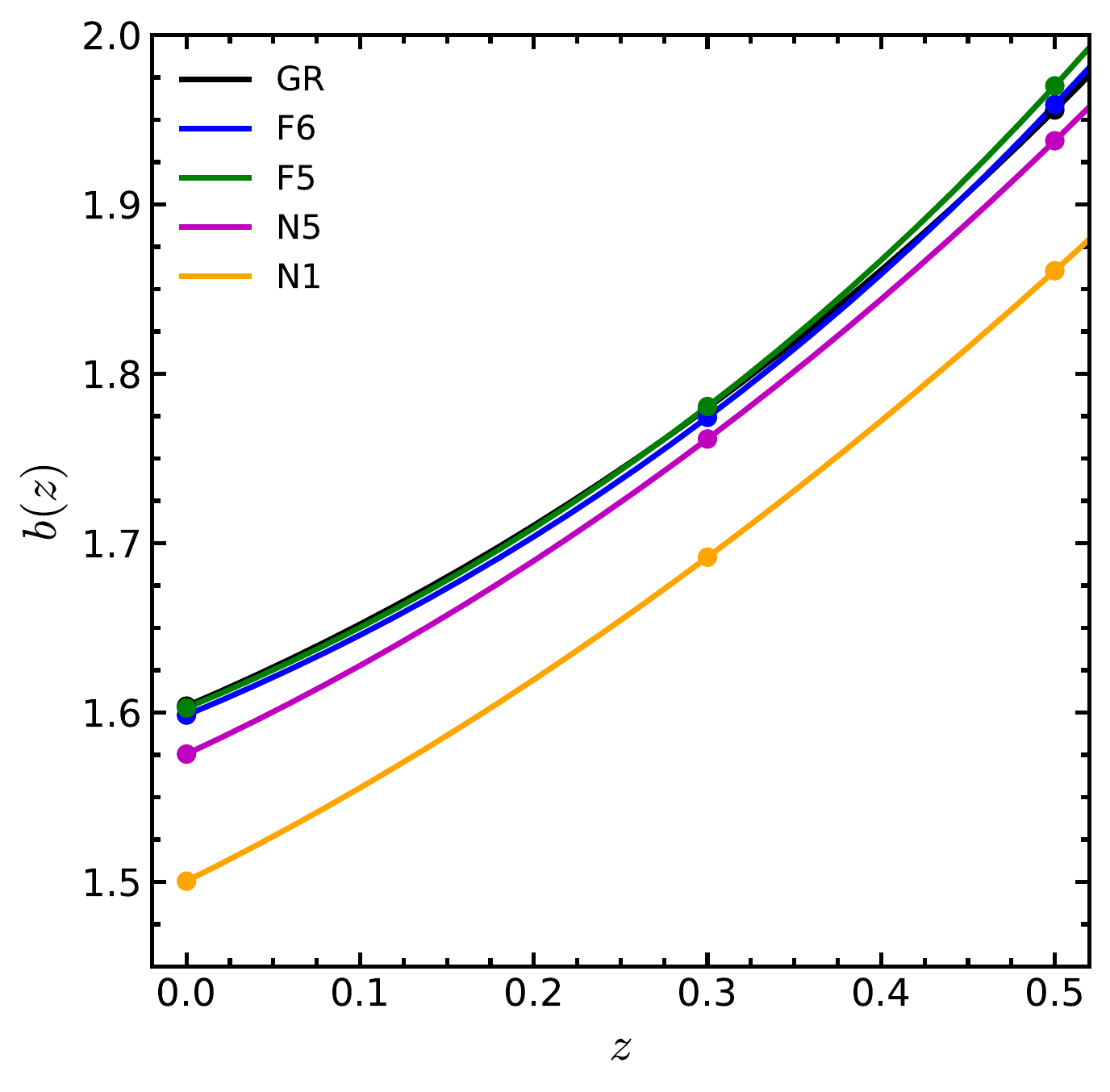}
	\caption{Linear galaxy bias measured from our galaxy mock catalogues at $z=0$, $z=0.3$ and $z=0.5$ (see Table \ref{tab:bias}) for the five gravity models: GR (black), F6 (blue), F5 (green), N5 (magenta) and N1 (orange). The solid lines represent an extrapolation between points at $z=0$, $0.3$ and $0.5$.}
	\label{fig:bias_z}
\end{figure}
\subsubsection{Linear galaxy bias}\label{sec:b_lin}
Galaxies are biased tracers of the dark matter density field, hence the relation between the distribution of galaxies and matter is given by the linear galaxy bias $b$ defined as
\begin{equation}
b \equiv \delta_{\rm g}/\delta\,,
\end{equation}
where $\delta_{\rm g}$ is the galaxy density contrast and $\delta$ is the density contrast of matter. In terms of the correlation function, the linear galaxy bias can be estimated in different ways,
\begin{equation}\label{eq:b_lin}
b(r) = \sqrt{\frac{\xi_{\rm gg}(r)}{\xi_{\rm mm}(r)}} = \frac{\xi_{\rm gg}(r)}{\xi_{\rm gm}(r)} = \frac{\xi_{\rm gm}(r)}{\xi_{\rm mm}(r)}\,,
\end{equation}
where $\xi_{\rm gg}(r)$ and $\xi_{\rm mm}(r)$ are respectively the galaxy-galaxy and matter-matter auto-correlation functions, and $\xi_{\rm gm}(r)$ is the galaxy-matter cross-correlation function, all in real space. \citet{Li:2017xdi} showed that the galaxy-matter cross-correlation coefficient, $R_{\rm gm}(r) \equiv \xi_{\rm gm}(r)/\sqrt{\xi_{\rm gg}(r)\xi_{\rm mm}(r)}$, approaches unity on scales $r \ge 2 \Mpch$, hence the linear bias measured in different ways agree well with each other. Therefore, we measure the linear galaxy bias from our mock catalogues as
\begin{equation}\label{eq:bias}
b(r,z) \equiv \frac{\xi_{\rm gg}(r,z)}{\xi_{\rm gm}(r,z)}\,,
\end{equation} 
which is less expensive to compute than $\xi_{\rm gm}/\xi_{\rm mm}$. At sufficiently large scales we expect $b(r) \approx {\rm const.}$, hence to measure the linear galaxy bias from our mock catalogues we make a fit of Eq.~\eqref{eq:bias} to a constant function using data in the range $r_{\rm min}\leq r\leq r_{\rm max}$, with $r_{\rm max} = 150 \Mpch$ and $10 < r_{\rm min}/(\Mpch) < 45$, then we take the mean over all best-fitting values. Fig.~\ref{fig:bias_z} shows the evolution of $b(r,z)$ as a function of $z$. 
We observe that, for the same number density, galaxies at higher redshifts are more biased tracers of the underlying dark matter field, which is due to faster growth of clustering of dark matter than of galaxies. Since the clustering amplitude of galaxies in real space is tuned to be the same for different cosmological models, by the construction of HOD, models with higher $\sigma_8$ (such as N1) have a lower linear bias as is shown in Fig.~\ref{fig:bias_z}.

The linear bias values measured from the mock galaxy catalogues for the different gravity models at the three different redshifts are listed in Table~\ref{tab:bias}.

\begin{table}
\centering
\caption{Linear galaxy bias, $b$, estimated by Eq.~\eqref{eq:bias} for all gravity models at $z=0$, $0.3$ and $0.5$.}
\label{tab:bias}
\begin{tabular}{cccc}
\hline
\multicolumn{4}{c}{Linear galaxy bias $b$} \\ \hline
      & $z=0$    & $z=0.3$    & $z=0.5$    \\ \hline
GR    & $1.6038$ & $1.7798$   & $1.9557$   \\
F6    & $1.5985$ & $1.7743$   & $1.9589$   \\
F5    & $1.6027$ & $1.7807$   & $1.9699$   \\
N5    & $1.5756$ & $1.7615$   & $1.9375$   \\
N1    & $1.5004$ & $1.6917$   & $1.8608$   \\ \hline
\end{tabular}
\end{table}
\section{Galaxy clustering in redshift space}\label{sec:RSD}
Peculiar velocities of galaxies induce anisotropies in redshift space and leave distinctive imprints on the clustering pattern at different regimes. On large (linear) scales, galaxies infall into high-density regions such as clusters producing a squashing effect of these regions along the line-of-sight: this is the Kaiser effect \citep{Kaiser:1987}. On smaller (nonlinear) scales, the random motions of galaxies in virialised objects produce the Fingers-of-God (FoG) effect where the density field becomes stretched and structures seem elongated along the line of sight \citep{FoG:Jackson}.
The amplitude of the RSD is related to the distortion parameter $\beta$, defined as
 \begin{equation}\label{eq:beta}
 \beta(z) \equiv \frac{f(z)}{b(z)}\,,
 \end{equation}
where $f$ is the linear growth rate (Eq.\eqref{eq:f_lin}) and $b$ is the linear galaxy bias (Eq.\eqref{eq:bias}) as a function of redshift. 
\begin{table}
\centering
\caption{True theoretical values of the $\beta$ parameter at $z=0$, $0.3$ and $0.5$ for the five gravity models. Since the growth rate is scale-dependent in $f(R)$ gravity we present the true values at 2 scales, $k = 0.01\hMpc$ and $k = 0.1\hMpc$.}
\label{tab:beta}
\begin{tabular}{cccc}
\hline
\multicolumn{4}{c}{$\beta_{\rm true}$} \\ \hline
    & $z=0$    & $z=0.3$  & $z=0.5$    \\ \hline
GR  & $0.3081$ & $0.3671$ & $0.3749$   \\
N5  & $0.3193$ & $0.3769$ & $0.3842$   \\
N1  & $0.3507$ & $0.4094$ & $0.4164$   \\ \hline
\multicolumn{4}{c}{$(k = 0.01 \hMpc)$} \\ \hline
F6  & $0.3091$ & $0.3682$ & $0.3744$   \\
F5  & $0.3087$ & $0.3672$ & $0.3725$   \\ \hline 
\multicolumn{4}{c}{$(k = 0.1 \hMpc)$}  \\ \hline
F6 & $0.3124$ & $0.3716$ & $0.3773$    \\
F5 & $0.3292$ & $0.3893$ & $0.3925$    \\ \hline
\end{tabular}
\end{table}

The fiducial value of $\beta$ for the five gravity models (GR, F6, F5, N5 and N1) at $z=0$, $0.3$ and $0.5$ are presented in Table~\ref{tab:beta}. Given the fact the linear growth rate, $f$, is scale-dependent in $f(R)$ gravity we present the fiducial values at two different wavenumbers ($k = 0.1\hMpc$ and $k = 0.01\hMpc$) corresponding to quasilinear and linear scales. The estimation of the linear bias parameter is taken from Table~\ref{tab:bias}.

\begin{figure}
 \centering
\includegraphics[width=0.5\textwidth]{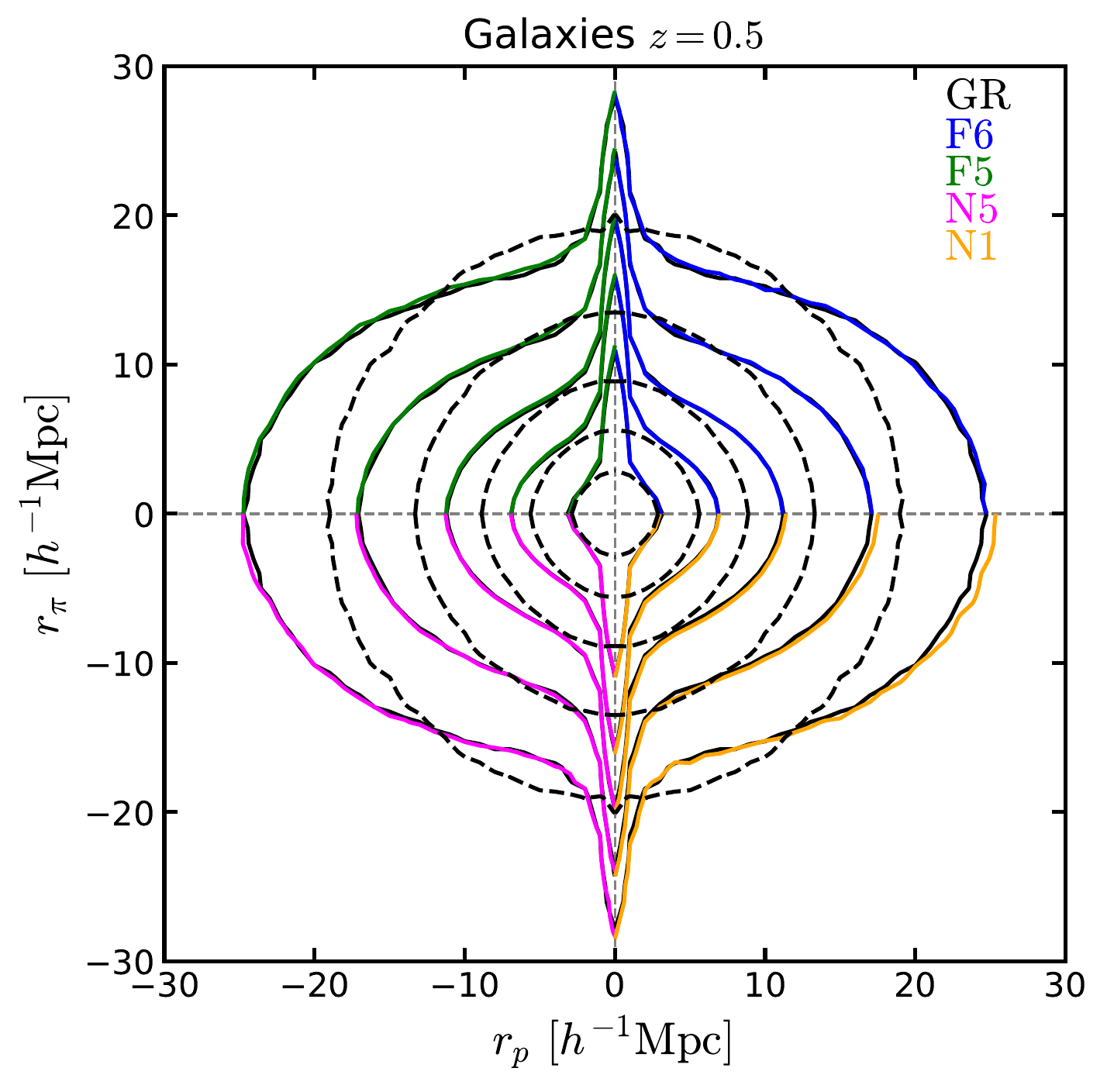}
	\caption{The two dimensional galaxy correlation function $\xi(r_p,r_\pi)$ measured from our mock catalogues at $z=0.5$ as a function of separation across $(r_p)$ and along $(r_\pi)$ the line-of-sight. Contours show lines of constant $\xi(r_p,r_\pi)$ at $\xi(r_p,r_\pi)$ = 5, 2, 1, 0.5, 0.25. The correlation functions correspond to the average of fifteen measurements obtained by projecting five realisations over the three LOS directions. For clarity we have projected $\xi(r_p,r_\pi)$ for GR for positive $r_p$ and $r_\pi$ over four quadrants and displayed the MG correlation function in different quadrants as follows: $(r_p,r_\pi)$ for F6, $(-r_p,r_\pi)$ for F5, $(-r_p,-r_\pi)$ for N5 and $(r_p,-r_\pi)$ for N1. The black dashed contours correspond to the real-space measurements of the correlation function at the same values of its counterpart in redshift-space, since all galaxy catalogues produce the same real-space clustering we show the GR case only. Different colour line correspond to different gravity model as specified in the legend.}
	\label{fig:2d_2pcf}
\end{figure}

The effects of redshift space distortions can be measured using the redshift-space correlation function of galaxies, $\xi(r_p,r_\pi)$, which is the excess probability of finding a pair of galaxies at separations transverse ($r_p$) and parallel ($r_\pi$) to the LOS. 
Fig.~\ref{fig:2d_2pcf} shows the redshift space correlation function $\xi(r_p,r_\pi)$ as a function of separation $(r_p,r_\pi)$ at $z=0.5$, for the different gravity models. 
The black dashed curve corresponds to the two-dimensional correlation function in real-space (since the clustering for different gravity models have been tuned to match each other, for demonstration, we just show the GR result).
We can clearly see that along the LOS at $r_p \lesssim 2\Mpch$ the clustering is enhanced by the peculiar velocities of galaxies producing the FoG effect, while at $r_p > 2\Mpch$ the clustering pattern is squashed thanks to the Kaiser effect. We observe that for nDGP models the contours become more flattened compared with GR because of the stronger linear growth rate (see Fig.~\ref{fig:f_D_z}). 
In the linear regime the overdensity grows proportional to the linear growth factor $\delta_{\rm m}(z) \propto D(z)$, therefore, the matter power spectra for the modified gravity models have higher amplitude and resulting in a higher matter fluctuation, $\sigma_{8}$, compared to GR.
A higher matter fluctuation produces an increase of galaxies that infall into high-density regions and makes the Kaiser effect stronger \citep{Tinker:2005na}. At the same time, we note that the FoG effect is very similar for the N1 and GR models, which is likely due to the effective Vainshtein screening mechanism on small scales in real space \citep[e.g.,][]{Paillas:2018wxs}, which makes the velocity dispersion comparable for these models.

In the case of $f(R)$ gravity models the two-dimensional correlation functions are indistinguishable from the one measured from the GR model. This is different from the behaviour of the nDGP models. A likely reason for this difference is the fact that in nDGP the fifth force is unscreened on large scales (i.e., beyond the Vainshtein radius of massive objects) so that the infall on scales of order $10\Mpch$ is enhanced, while for the $f(R)$ models considered here the fifth force is short ranged and cannot affect such scales.

In order to increase the signal-to-noise ratio, it is helpful to further project the two-dimensional correlation function $\xi(r_p,r_\pi)$ onto a one-dimensional object. Given the symmetry along the line-of-sight, we first express the transverse and parallel separation $(r_p,r_\pi)$ as separation in redshift space and the cosine of the angle between $s$ and the LOS direction, 
\begin{equation}
s = \sqrt{r^2_\pi + r^2_p}\,, \qquad \mu = \frac{r_\pi}{s}\,.
\end{equation}

\begin{figure*}
 \centering
\includegraphics[width=0.45\textwidth]{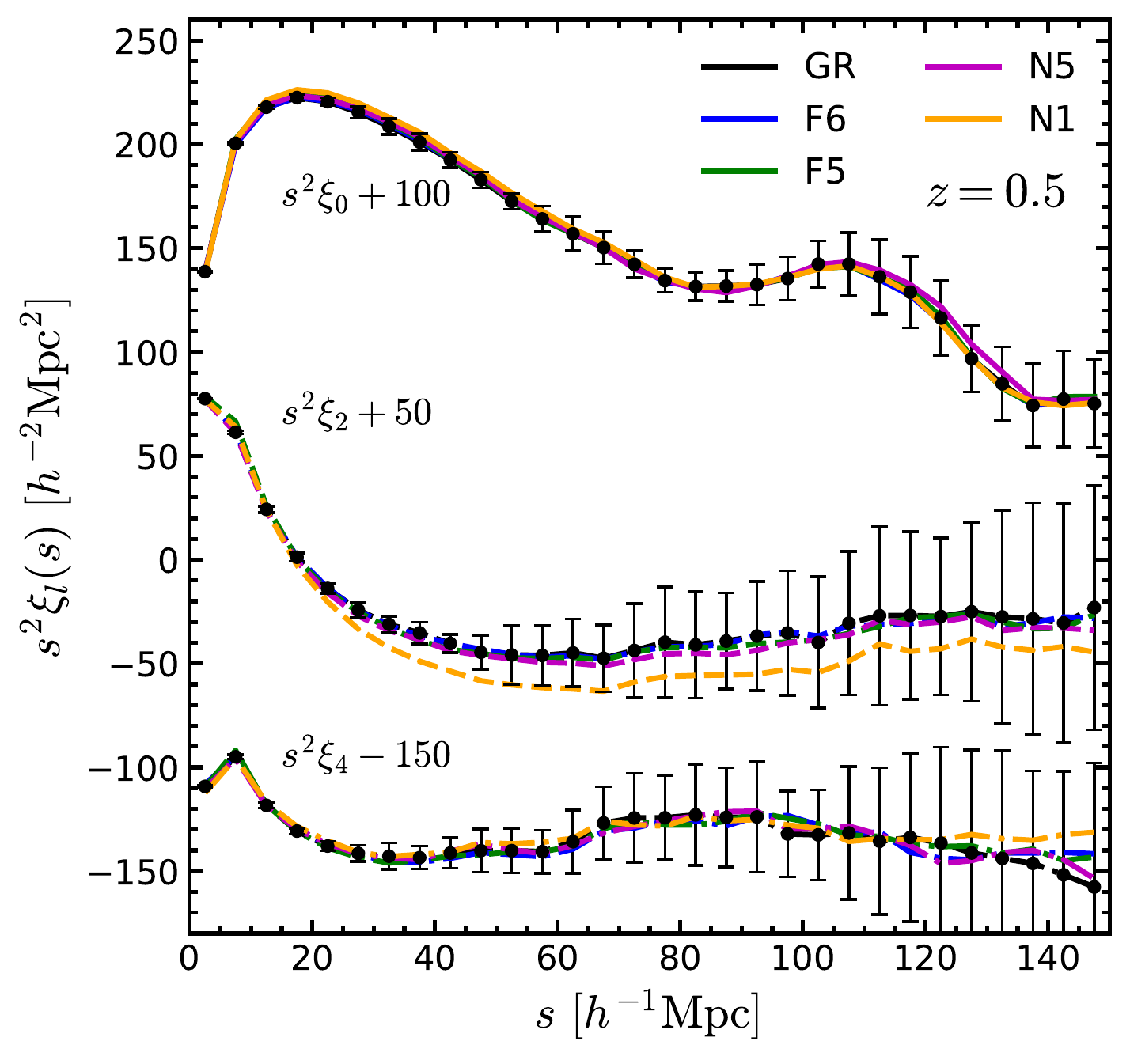}
\includegraphics[width=0.45\textwidth]{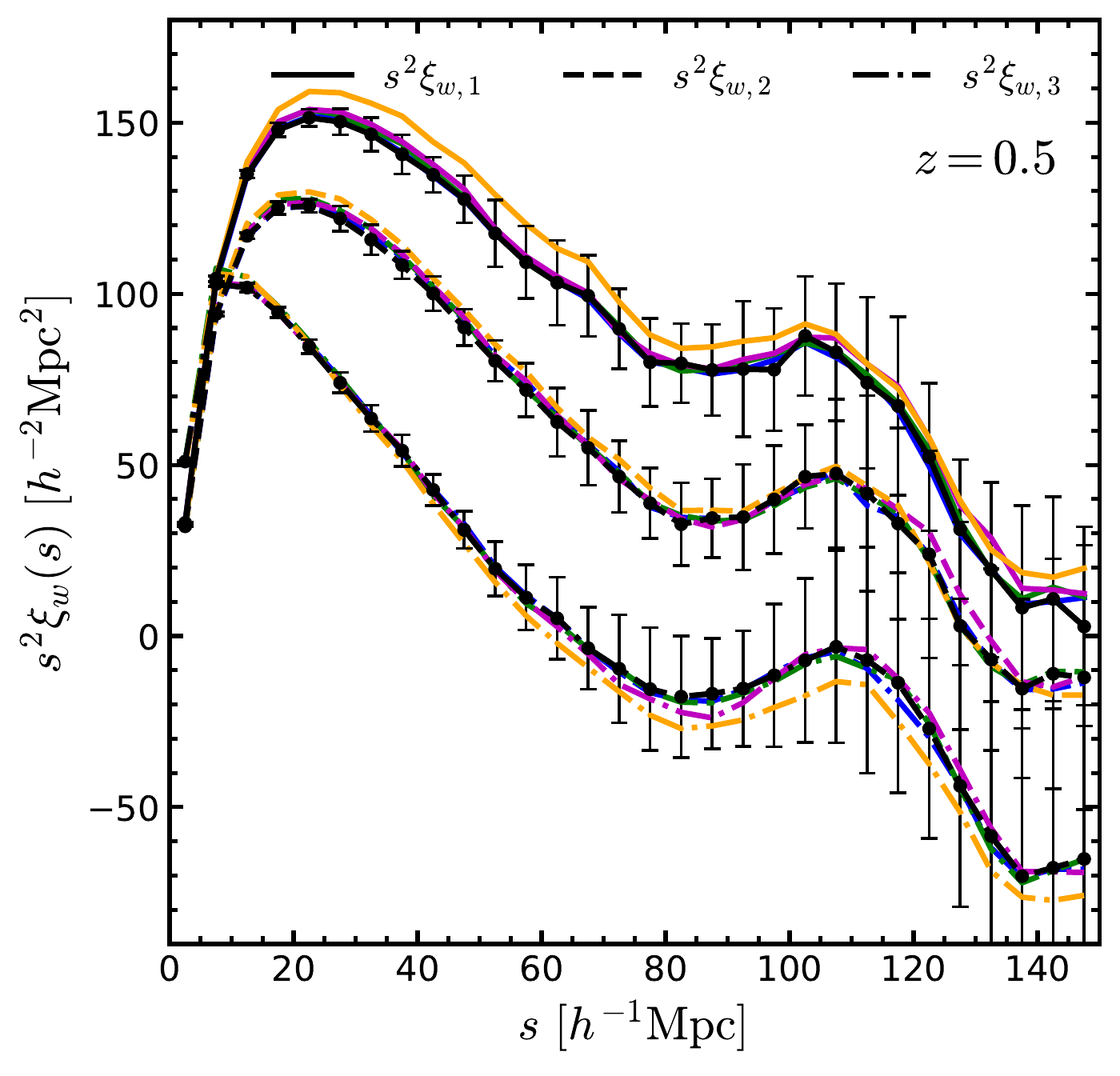}
	\caption{{\it Left panel}: Monopole, quadrupole and hexadecapole moments of the correlation function, Eq.~\eqref{eq:xi_l}, for our five gravity models at $z=0.5$. The moments have been shifted by a factor of $100$, $50$ and $-150$ for better visualization. {\it Right panel}: Clustering wedges, Eq.~\eqref{eq:xi_w}, measured at $z=0.5$ for all gravity models as labelled, the upper wedge (solid lines) correspond to angles with $0 < \mu < 1/3$, the middle wedge (dashed lines) to $1/3 < \mu < 2/3$ and the lower wedge (dot-dashed lines) to $2/3 < \mu < 1$. The error bars correspond to the standard deviation over fifteen GR measurements.}
	\label{fig:xi_lw}
\end{figure*}
We decompose $\xi(s,\mu)$ into multipole moments,
\begin{equation}\label{eq:xi_l}
\xi_l(s) = (2l + 1) \int^1_{0}{\xi(s,\mu)} P_l(\mu)~{\rm d}\mu
\end{equation}
where $P_l(\mu)$ are the Legendre polynomials. In the linear regime, the $l=0$, $2$ and $4$ moments are non-zero with $P_0(\mu) = 1,\, P_2(\mu) = (3\mu^2 + 1)/2,\, P_4(\mu) = (35\mu^4 - 30\mu^2 +3)/8$, corresponding to the monopole, quadrupole and hexadecapole moments. We measured $\xi(s,\mu)$ from our galaxy catalogues using linear bins centred at 2.5 to 147.5 $\Mpch$ with separation $\Delta s = 5 \Mpch$. For $\mu$, we use 30 linearly spaced bins between $0$ and $1$.

The left panel of Fig.~\ref{fig:xi_lw} shows the multipole moments ($\xi_l(s)$) of the correlation functions measured from our galaxy catalogues at $z=0.5$ for the different gravity models. From the monopole, $\xi_0(s)$ (upper curves of left panel in Fig.~\ref{fig:xi_lw}), we observe that the position of the baryon acoustic oscillation (BAO) peak is not affected by modified gravity and can be found at a scale of $s_{\rm BAO} \simeq 105\Mpch$ or $150~{\rm Mpc}$. 
Higher  order multipole moments such as the quadrupole ($\xi_2(s)$) and the hexadecapole ($\xi_4(s)$) encode the degree of anisotropy produced by redshift distortions. 
In the case of the quadrupole, $\xi_2(s)$, N1 shows the strongest deviation with respect to GR especially on scales $s>20\Mpch$, followed by N5. This is a direct consequence of a more squashed contour for the nDGP models caused by higher growth rate and stronger matter fluctuation.
Our measurements of the hexadecapole are almost indistinguishable when compare the MG models with GR. This is due to the fact that higher order multipoles $(l \geq 4)$ do not have a big impact on the estimation of the correlation function and are noisier than the monopole and quadrupole \citep{Hamilton:1997zq}.

As an alternative to the multipoles, we also measured the clustering wedges which correspond to angle-averaged measurements of the correlation function \citep{Kazin:2012wedges},
\begin{equation}\label{eq:xi_w}
\xi_w(s) = \frac{1}{\mu_2 - \mu_1} \int^{\mu_2}_{\mu_1}{\xi(s,\mu)}~{\rm d}\mu.
\end{equation}
In this work we choose the intervals $(i-1)/3 < \mu < i/3$ with $i=1,2,3$, which are commonly used in the literature (for instance, see \citealt{Sanchez:2016sas}). 
The relation between multipoles and wedges is given by the transformation,
\begin{equation}\label{eq:xiw_l}
\xi_w(s) = \sum_l \xi_l(s)\bar{P}_l\,,
\end{equation}
where $\bar{P}_l$ is the average of the Legendre polynomial over the $\mu$-bin.
When the higher order statistics can be truly neglected, an explicit expression can be written down as the transformation between the multipoles and wedges,
\begin{equation}\label{eq:wedge-transform}
\xi_{\ell}(s)=
  \left[ {\begin{array}{ccc}
   1/3 & 1/3 & 1/3 \\
   -9/14 & -15/28 & 33/28 \\
   54/35 & -81/35 & 27/35\\
  \end{array} } \right] \xi_w(s)\,.
\end{equation}
Where the mean of the three clustering wedges correspond to the monopole of the correlation function,
\begin{equation}\label{eq:xiw_0}
\xi_0(s) = \frac{\xi_{w,1}(s) + \xi_{w,2}(s) + \xi_{w,3}(s)}{3}\,.
\end{equation}

The measured clustering wedges ($\xi_w(s)$) at $z=0.5$ are shown in the right panel of Fig.~\ref{fig:xi_lw}. The behaviour of the 2D two-point correlation functions, $\xi(r_p,r_\pi)$, shown in Fig.~\ref{fig:2d_2pcf}, can be quantified and described by the clustering wedges. The first wedge, $\xi_{w,1}$, encodes information about the correlation function closer to the transverse (or perpendicular) direction ($r_p$). Here, the FoG effect is not significant and the Kaiser effect governs the clustering of galaxies. For this reason (see the description of $\xi(r_p,r_\pi)$ above) N1 has a larger positive amplitude compared to GR, followed by N5 but with a weaker deviation. The second or intermediate wedge, $\xi_{w,2}$, corresponds to $\bar{\mu} = 0.5$ and is the closest to the monopole in shape and amplitude. Finally, the third wedge, $\xi_{w,3}$, which is closer (or parallel) to the LOS ($r_\pi$), is the most impacted by the random motions of galaxies due to the FoG effect. The shape of the parallel wedge is slightly different to the transverse and intermediate wedges, and on smaller scales it has a steeper slope. In this case, we found a negative difference between the MG models with respect to GR (opposite to the difference found in the transverse wedge).

In general, both the multipoles and the wedges of the correlation function for $f(R)$ gravity models show a weaker deviation from GR and this is expected to impact on the estimation of $\beta$. In the following sections we discuss how to estimate $\beta$ from theoretical models based on perturbation theory.

\begin{figure*}
 \centering
\includegraphics[width=0.45\textwidth]{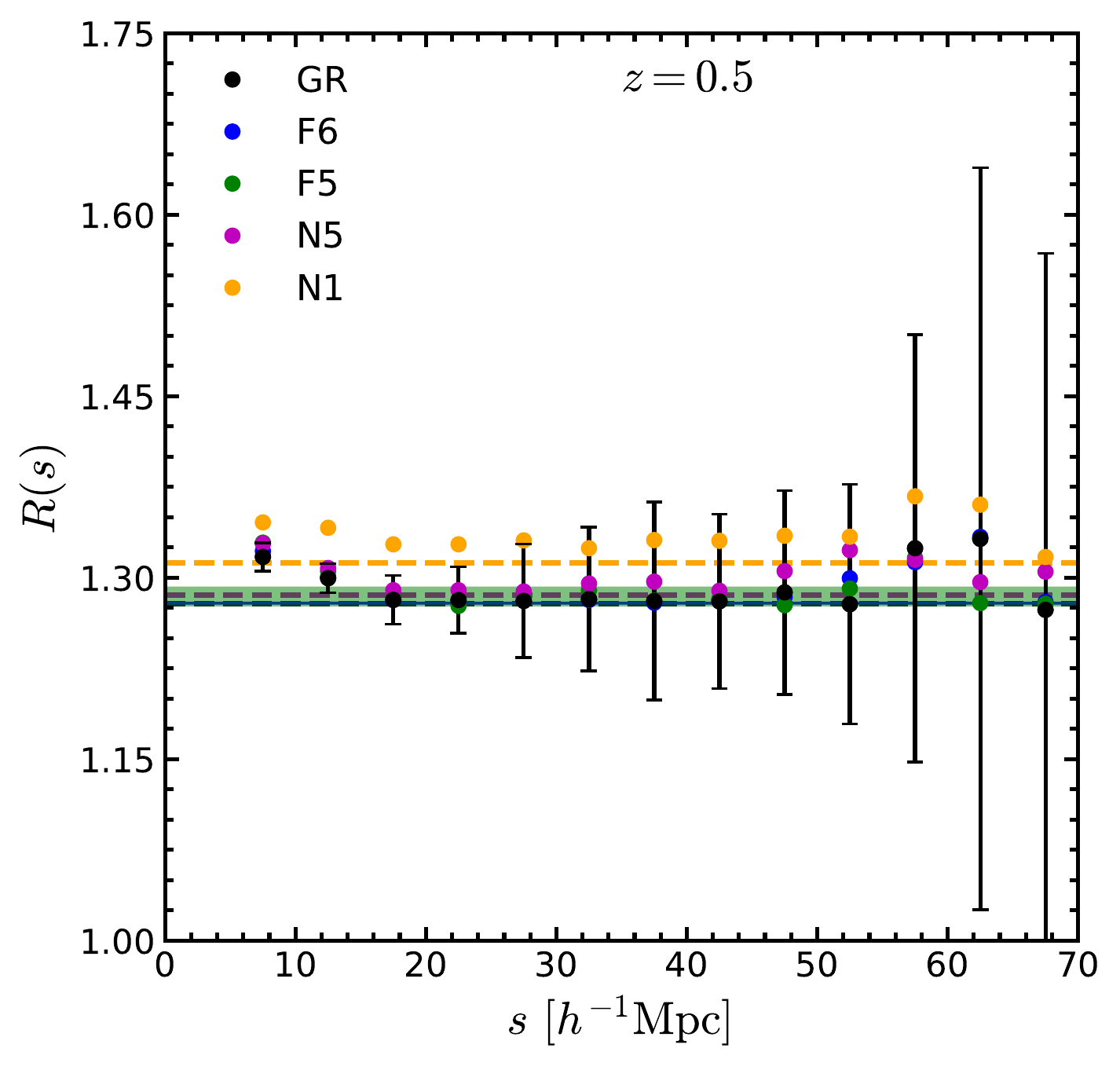}
\includegraphics[width=0.45\textwidth]{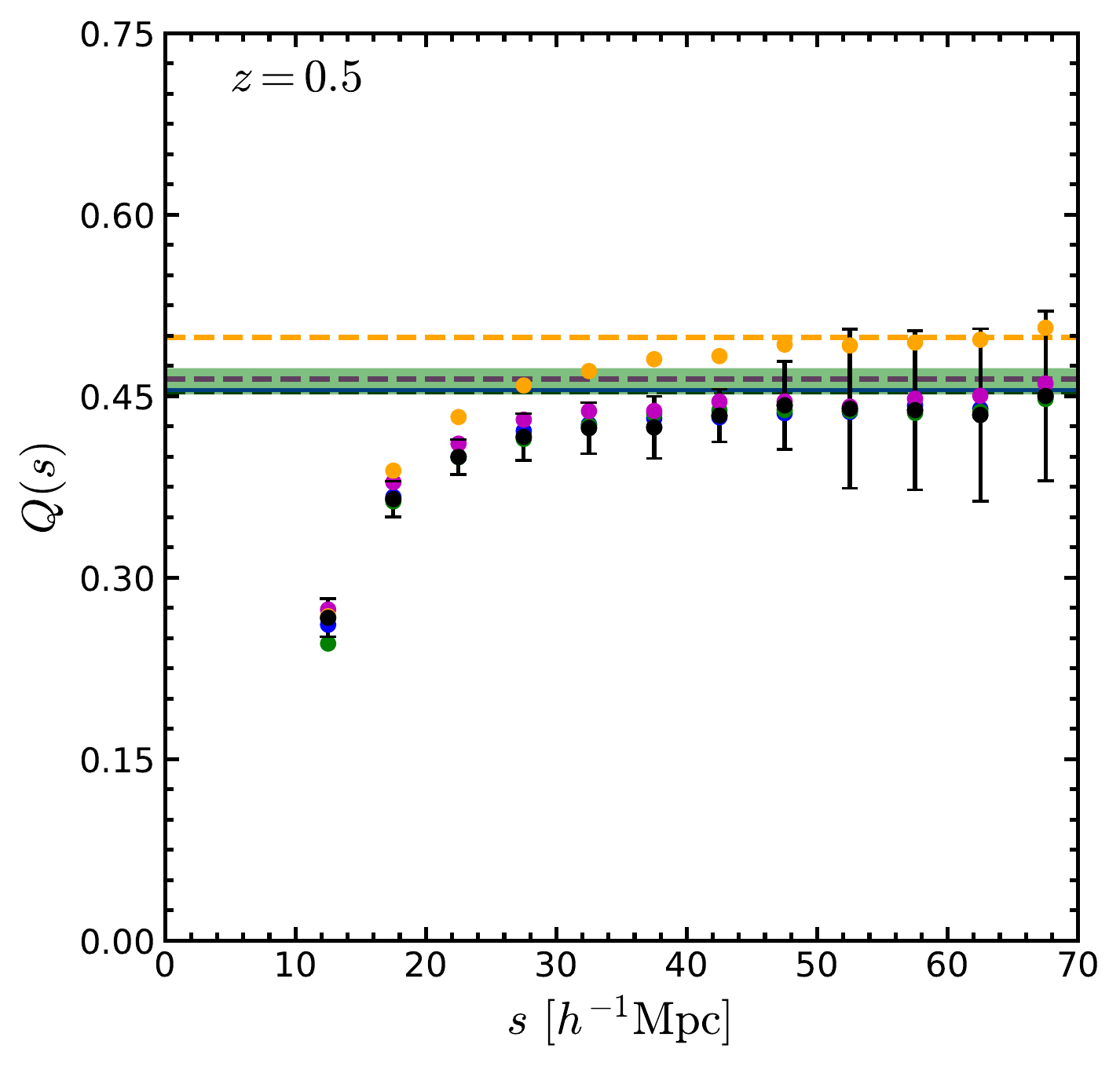}	
	\caption{{\it Left panel}: Ratio of the monopole in redshift space to the real space correlation function, $R(s)$ Eq.~\eqref{eq:Rs}, as a function of separation at $z=0.5$. {\it Right panel}: $Q(s)$ estimator defined by Eq.\eqref{eq:Qs} as a function of separation at $z=0.5$. In both panels we plot the estimators up to a scale of $s=70\Mpch$ for better visualisation and to avoid the transition to negative values of $\xi_0(s)$. Horizontal dashed lines represent the theoretical predictions of the linear model for GR, N5 and N1 models, for the case of $f(R)$ gravity the theoretical predictions are shown as horizontal shaded bands. The error bars correspond to the standard deviation over fifteen GR measurements.}
    \label{fig:RQ_sim}
\end{figure*}

\section{Theoretical RSD models and parameter estimation}
\label{sect:model_constraints}

In this section we give the main results of this paper, namely the validation of the inference of $\beta$ based on a number of estimators of RSD, using the redshift-space mock galaxy catalogues mentioned above. For a given set of model parameters (e.g., $\beta$), the theoretical predictions of the estimators are obtained using two methods -- linear perturbation theory (Kaiser model) and the Galilean-invariant renormalized perturbation theory (gRPT) based on higher-order perturbation theory. We shall discuss these two methods and their results in two separate subsections, and discuss the implications of the results in a third subsection.

\subsection{Linear model}\label{sec:Kaiser}

In linear perturbation theory, the relation between the redshift-space galaxy power spectrum with its counterpart in real space is given by the Kaiser formula \citep{Kaiser:1987}: 
\begin{equation}\label{eq:Pk_s_lin}
P_{\rm s}(k,\mu) = (1 + \beta \mu^2)^2 P_{\rm r}(k)\,.
\end{equation}
As we are interested in the effects of RSD on the correlation function, we need to have a similar relation in configuration space. 
Under the plane parallel approximation of the distortion operator, the correlation function is expressed as follows \citep{Hamilton:1992zz,Hamilton:1997zq},
\begin{eqnarray}
\xi(s,\mu) &=& [1 + \beta (\partial/\partial z)^2(\nabla^2)^{-1}]^2\xi(r)\,,\\
~&=& \xi_0(s)P_0(\mu) + \xi_2(s)P_2(\mu) + \xi_4(s)P_4(\mu)\,.\label{eq:xi_s_lin}
\end{eqnarray}
In linear theory, the multipoles of the correlation function can be estimated as follows \citep{Hamilton:1992zz},
\begin{eqnarray}
\xi_0(s) &=& \left( 1 + \frac{2\beta}{3} + \frac{\beta^2}{5} \right)\xi(r)\,,\label{eq:xi0_lin}\\
\xi_2(s) &=& \left( \frac{4\beta}{3} + \frac{4\beta^2}{7} \right)[\xi(r) - \bar{\xi}(r)]\,,\label{eq:xi2_lin}\\
\xi_4(s) &=& \frac{8\beta^2}{35}\left[\xi(r) + \frac{5}{2}\bar{\xi}(r) - \frac{7}{2} \bar{\bar{\xi}}(r) \right]\,,\label{eq:xi4_lin}
\end{eqnarray}
where $\xi(r)$ is the galaxy correlation function in real-space and
\begin{eqnarray}
\bar{\xi}(r) &=& \frac{3}{r^3} \int^r_0 \xi(r')r'^2~{\rm d}r' \,,\\
\bar{\bar{\xi}}(r) &=& \frac{5}{r^5} \int^r_0 \xi(r')r'^4~{\rm d}r' \,.
\end{eqnarray}
From Eqs.~\eqref{eq:xi0_lin} and \eqref{eq:xi2_lin} we can define two estimators to obtain the distortion parameter, $\beta$ \citep{Hawkins:2002sg},
\begin{equation}\label{eq:Rs}
R(s) \equiv \frac{\xi_0(s)}{\xi(r)} = 1 + \frac{2\beta}{3} + \frac{\beta^2}{5}\,,
\end{equation}
and
\begin{equation}\label{eq:Qs}
Q(s) \equiv \frac{\xi_2(s)}{\xi_0(s) - \bar{\xi}_0(s)} = \frac{(4/3)\beta + (4/7)\beta^2}{1 + (2/3)\beta + (1/5)\beta^2}\,,
\end{equation}
where
\begin{equation}
\bar{\xi}_0(s) = \frac{3}{s^3} \int^s_0 \xi_0(s')s'^{2}~{\rm d}s'\,,
\end{equation}
is the volume average of the monopole in redshift space.
For the estimation of clustering wedges in the linear theory model we simply substitute Eqs.~\eqref{eq:xi0_lin}-\eqref{eq:xi4_lin} into Eq.~\eqref{eq:xiw_l}.
 
Figure~\ref{fig:RQ_sim} compares the theoretical and measured values of the two estimators -- $R(s)$ on the left and $Q(s)$ on the right. In both panels the dashed horizontal lines represent the theoretical predictions for GR and nDGP models of $R(s)$ and $Q(s)$; for $f(R)$ gravity models the theoretical predictions are shown as horizontal shaded bands for wavenumbers $0.01 < k/(\hMpc) < 0.1$. The theoretical estimations of $R(s)$ and $Q(s)$ are calculated with the second equality of Eqs.~\eqref{eq:Rs} and \eqref{eq:Qs}, respectively, where the values of $\beta$ are taken from Table~\ref{tab:beta}.

The symbols in the left panel of Fig.~\ref{fig:RQ_sim} show the measurements of $R(s)$ (first equality of Eq.~\eqref{eq:Rs}) from our galaxy catalogues at $z=0.5$ for the different gravity models. 
Let us recall that the real space correlation functions have been tuned to be within $2$-$3\%$ for all gravity models (see Section~\ref{sec:HOD}). Hence, the difference in $R(s)$ between models is mainly caused by the difference in the redshift space monopole, $\xi_0(s)$.
From the measurements from the simulations, we find that all models reach an asymptotic value at $s \approx 10\Mpch$ as expected \citep{Tinker:2005na,Marulli:2015jga}. We also find that for all models the mean values of $R(s)$ are above the theoretical expectation. The reason for which is that the Kaiser model, Eq.~\eqref{eq:Pk_s_lin}, does not contain a FoG term which models the power of galaxies on small separations and hence underestimates the clustering on all scales when Fourier transforming to get the correlation function. Nevertheless, given the size of the error bars, all models show a good agreement with the fiducial values.

The $Q(s)$ estimator at $z=0.5$ is presented in the right panel of Fig.~\ref{fig:RQ_sim}. The measurements are obtained by taking the ratio in the first equality of Eq.~\eqref{eq:Qs}. In this case the mean values from all models reach an asymptotic value at separations $s = 40\Mpch$ but are still below the fiducial values (opposite to $R(s)$), only matching with the theoretical expectation at $s \sim 70\Mpch$. On scales below $s = 30\Mpch$ non-linearities produce smaller values of $Q(s)$. Our results are consistent with previous observational and theoretical findings (see e.g. \citealt{Hawkins:2002sg,Tinker:2005na,Tinker:2006dm}).

\begin{figure*}
 \centering
\includegraphics[width=0.4\textwidth]{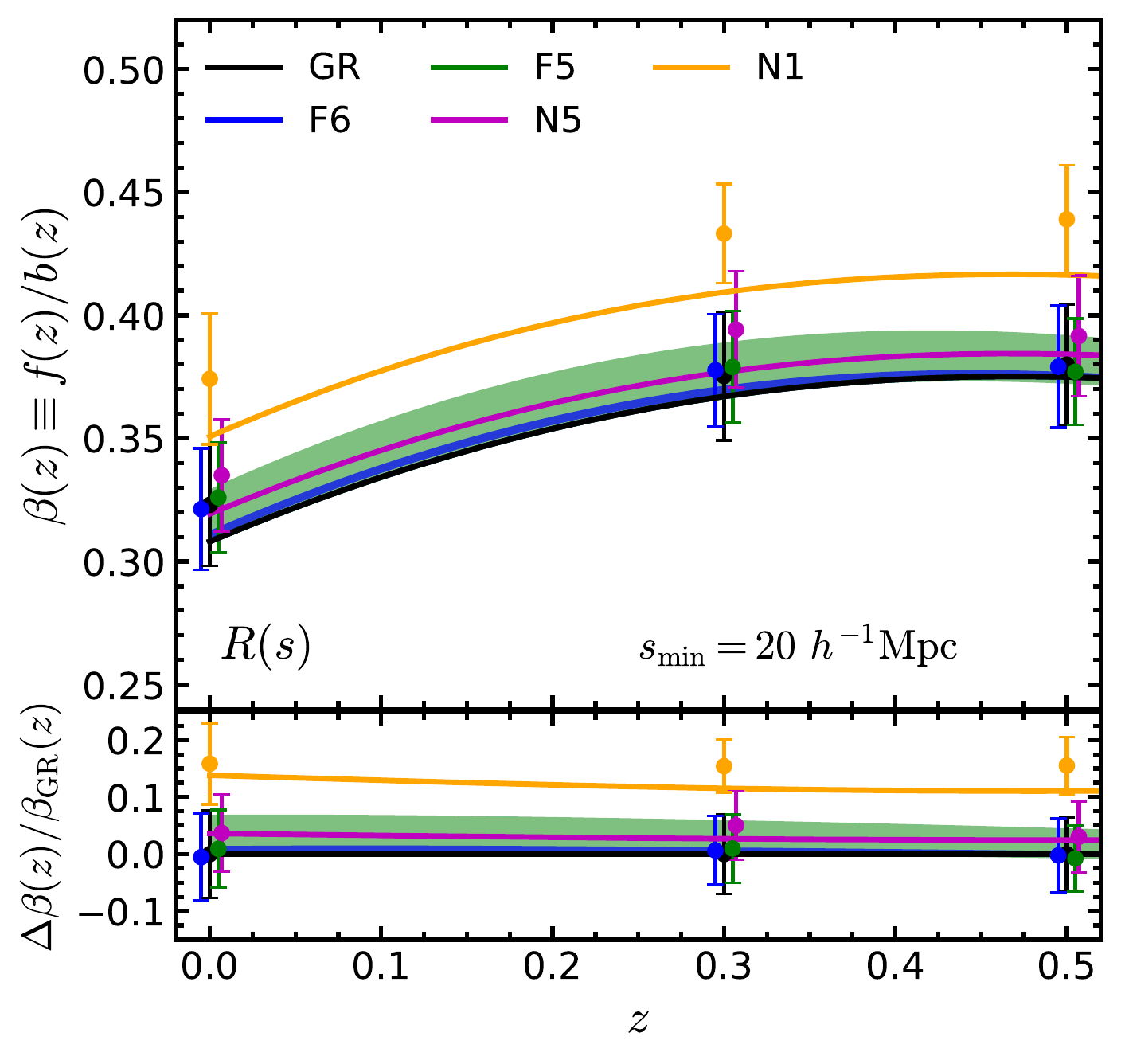}
\includegraphics[width=0.4\textwidth]{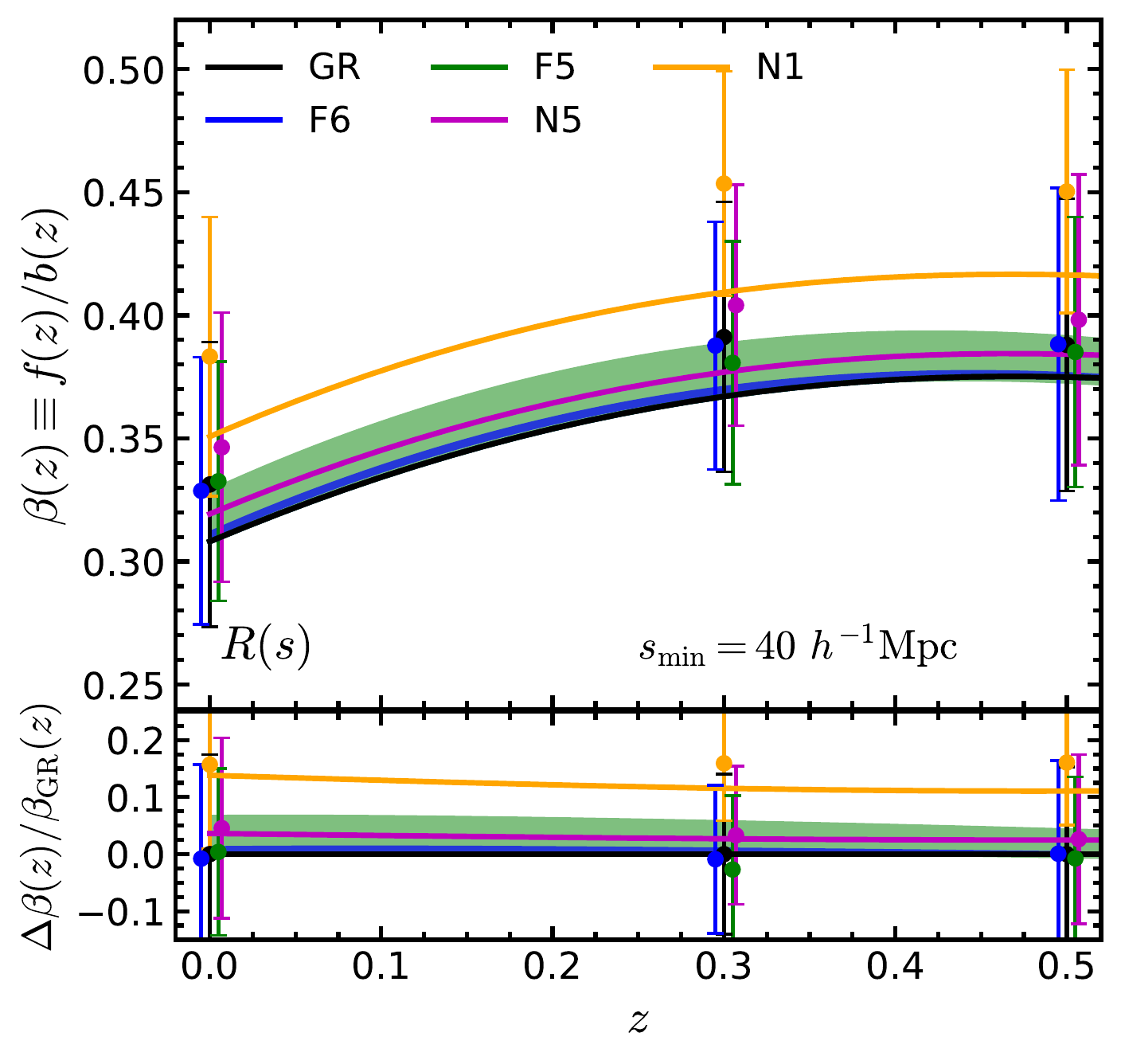}
\includegraphics[width=0.4\textwidth]{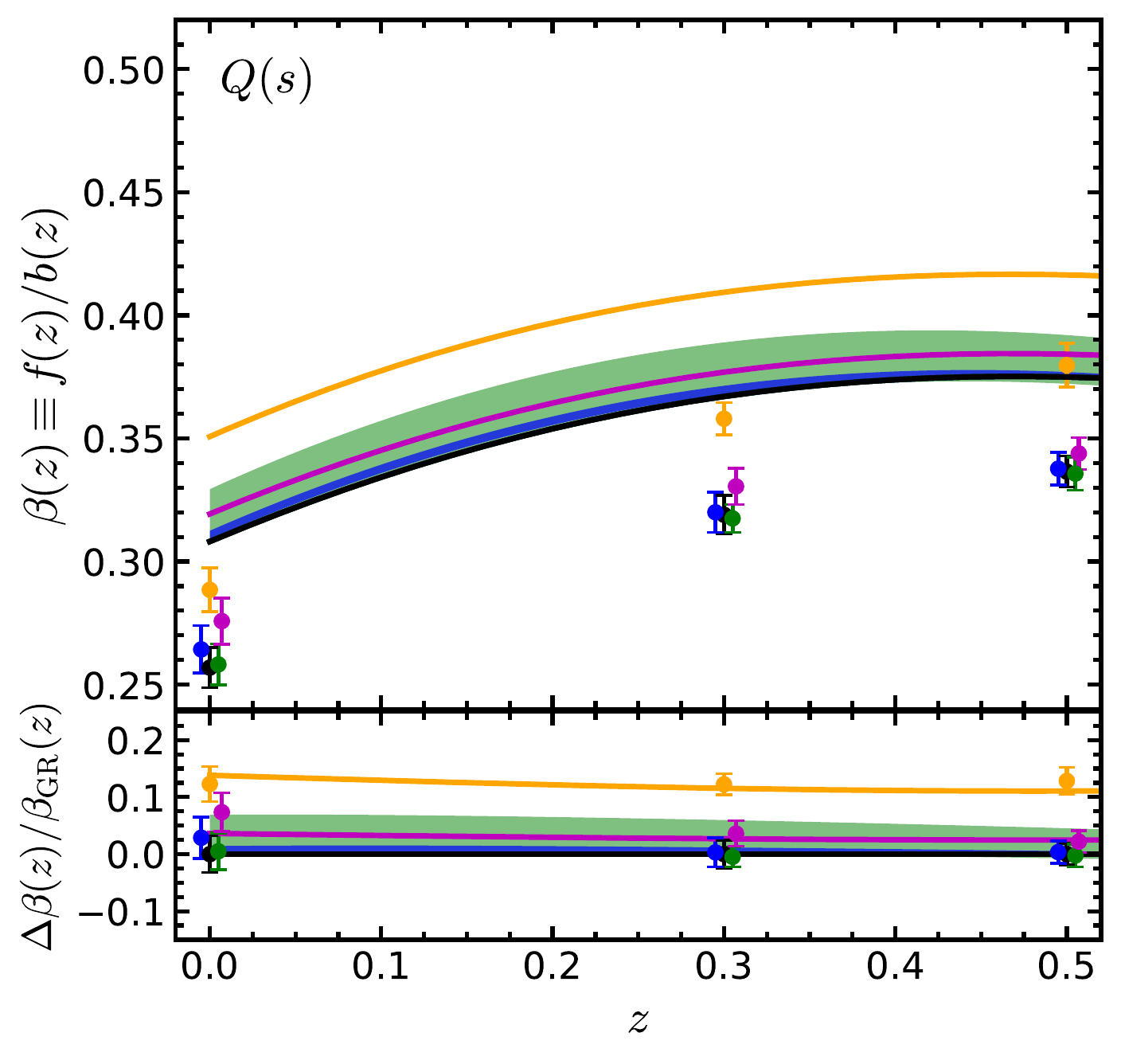}
\includegraphics[width=0.4\textwidth]{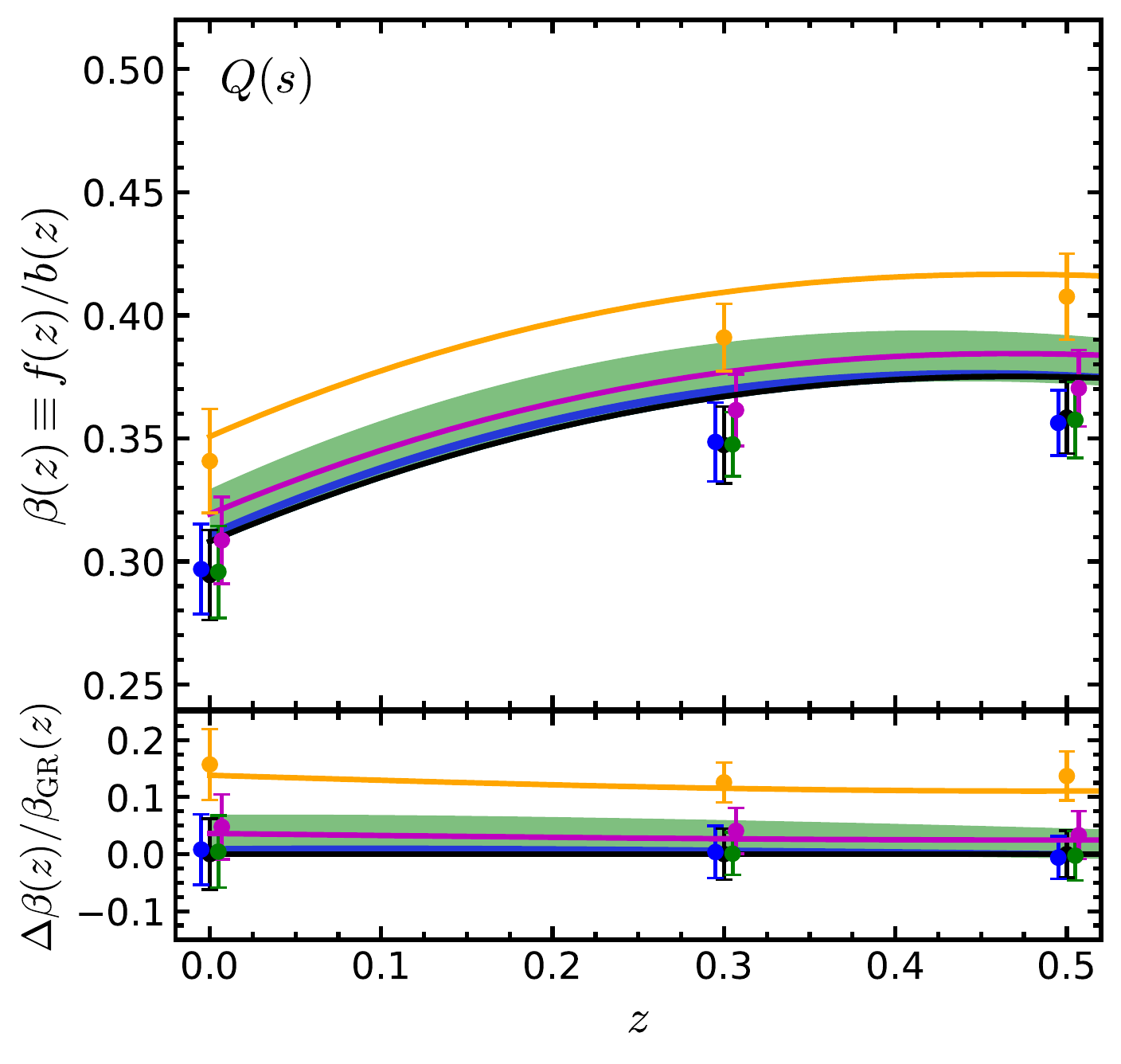}
\includegraphics[width=0.4\textwidth]{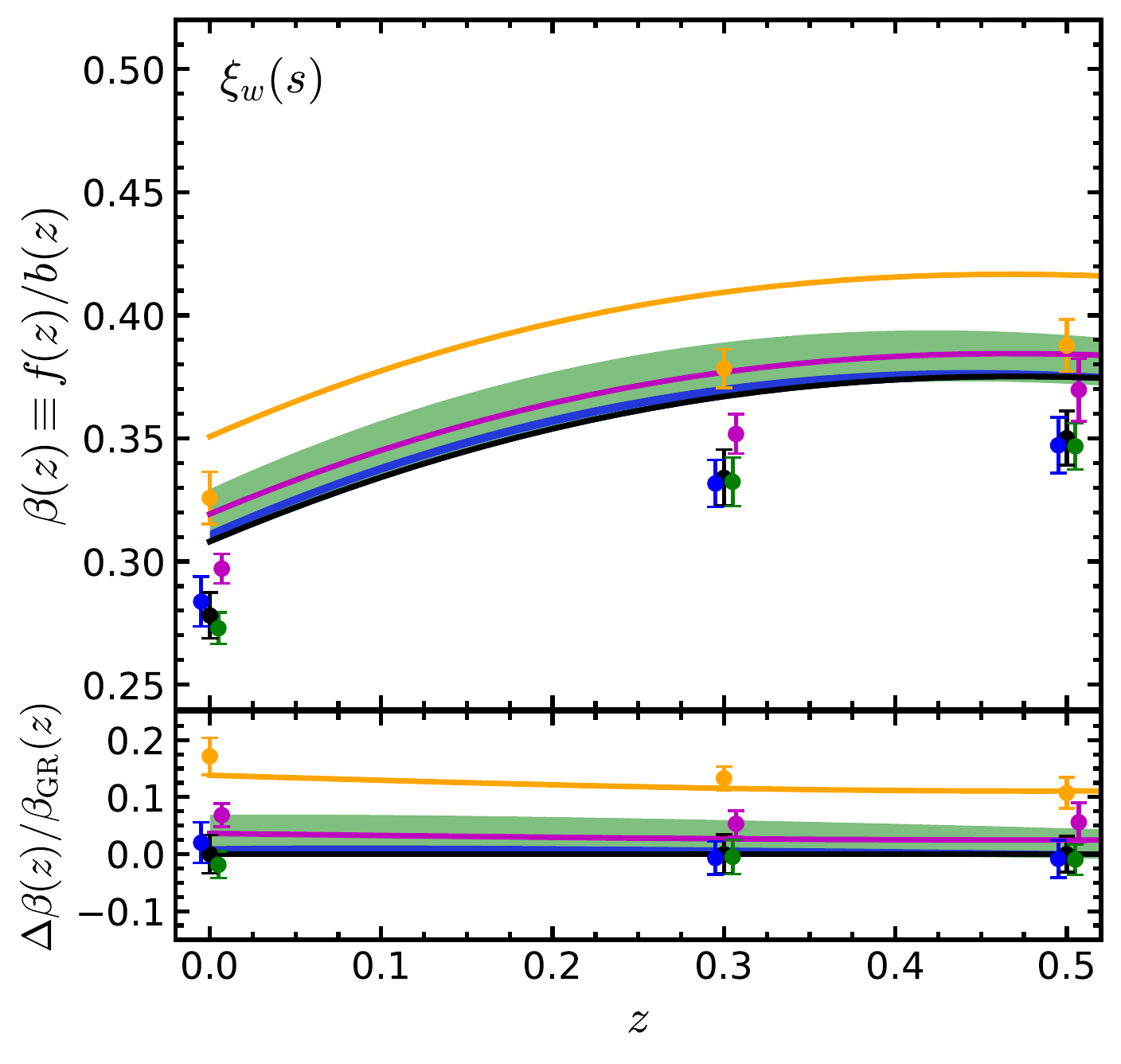}
\includegraphics[width=0.4\textwidth]{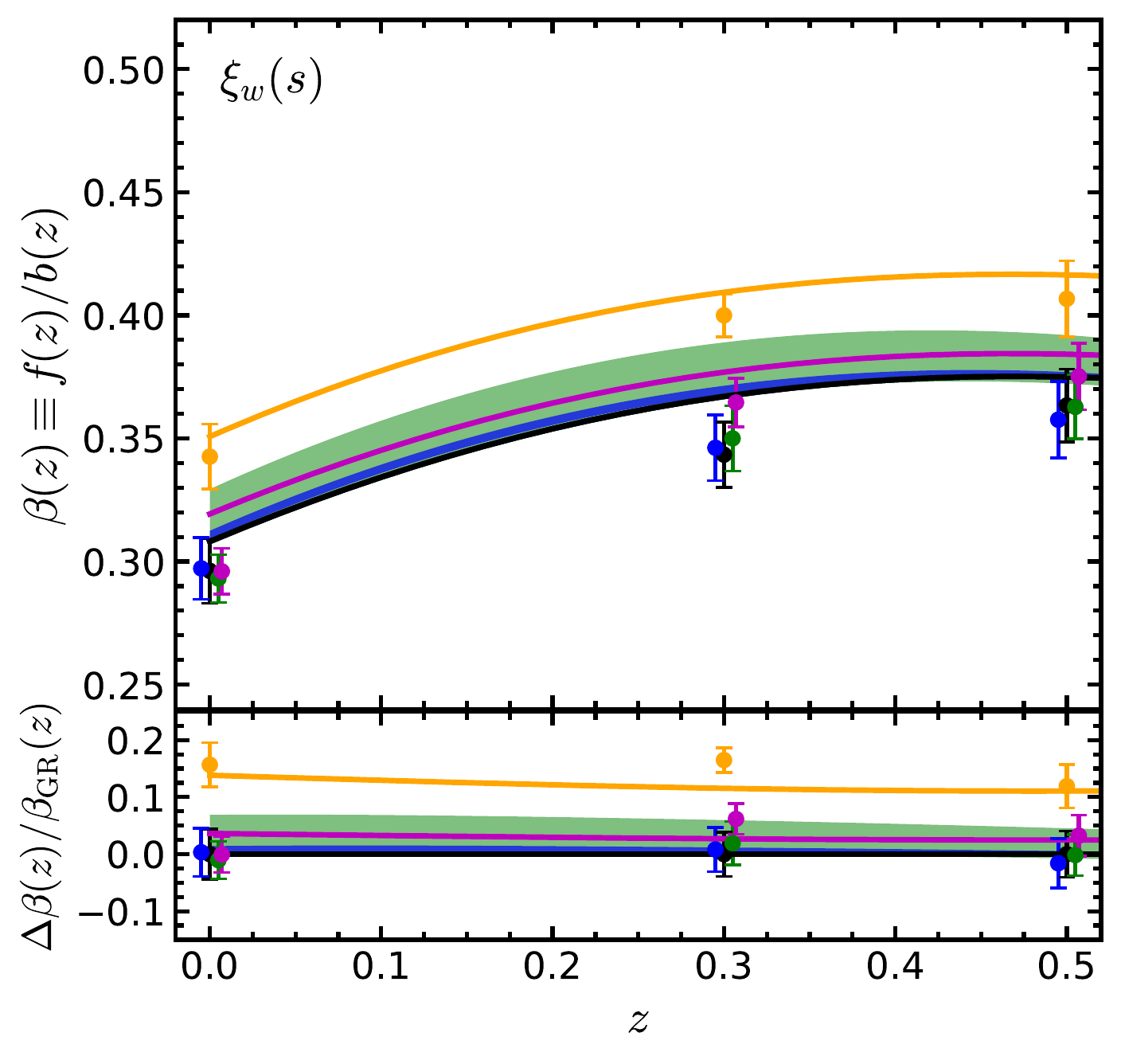}
	\caption{Evolution of $\beta$ as a function of redshift. Solid curves show the theoretical predictions for the gravity models as shown in the legend, for $f(R)$ gravity models the theoretical predictions are shown as a shaded region for wavenumbers $0.01 \leq k/[\hMpc] \leq 0.1$. Each panel shows the best-fitting $\beta$ values (filled symbols) using the estimators: $R(s)$ Eq.~\ref{eq:Rs} (upper panels), $Q(s)$ Eq.~\eqref{eq:Qs} (middle panels) and $\xi_w(s)$ Eq.~\eqref{eq:xiw_l} (lower panels) with $s_{\rm min} = 20\Mpch$ (left panels) and $s_{\rm min} = 40\Mpch$ (right panels). The lower subpanels show the relative difference between the modified gravity models and GR. Error bars correspond to the $1\sigma$ confidence level.} 
	\label{fig:beta_lin_s1}
\end{figure*}

\subsubsection{Parameter estimation using the linear model}
\label{sec:params}
To estimate $\beta(z)$ from $R(s)$, $Q(s)$ and $\xi_w(s)$ using the linear theory model, we use a $\chi^2$-test by minimising the $\chi^2$ defined as
\begin{equation}\label{eq:chi2}
\chi^2(\beta) = \sum_i \left( \frac{E(s_i) - E^{\rm{th}}(s_i;\beta)}{\sigma_{E_i}} \right)^2\,,
\end{equation}
where $E(s)$ is the average measured linear estimator (i.e. $R(s)$, $Q(s)$ or $\xi_w(s)$), $\sigma_{E}$ is the standard deviation over fifteen measurements and $E^{\rm{th}}(s;\beta)$ is the theoretical prediction of each estimator. 

To obtain the best-fitting value of $\beta$, we searched in a grid of values in the range $\beta \in [0,1]$, with a step size of $\Delta\beta = 0.0001$, for the theoretical estimators and identified the value of $\beta$ that minimises the $\chi^2$, Eq.~\eqref{eq:chi2}, as $\chi^2_{\rm min} = \chi^2(\beta_{\rm fit})$. Since we vary only one parameter, the $1\sigma$ error bar on $\beta$ corresponds to $\Delta\chi^2 \equiv \chi^2 - \chi^2_{\rm min} = 1$. 
We fit our measurements using two ranges of scales: $s = 20 - 150 \Mpch$ ($s_{\rm min}=20\Mpch$) and $s = 40 - 150 \Mpch$ ($s_{\rm min}=40\Mpch$).

In Fig.~\ref{fig:beta_lin_s1} we show the best-fitting $\beta$ values for all gravity models at $z=0$, $0.3$ and $0.5$ for the linear (Kaiser) model, with $1\sigma$ error bars, compared to with their theoretical prediction. The left column corresponds to the fits using the range $s = 20 - 150 \Mpch$ and the right column shows the fits for the range $s = 40 - 150 \Mpch$. The first row corresponds to the monopole to real space correlation function ratio, $R(s)$, the second row shows the best-fit values from using the $Q(s)$ estimator and the third row presents our results from using three clustering wedges, $\xi_w(s)$.

We find that the best-fitting $\beta$ values using $R(s)$ are above the theoretical expectations, in particular for the nDGP models, which is not surprising if we look at the left panel of Fig.~\ref{fig:RQ_sim} and note that our measurements for these models show an offset compared to the theoretical predictions. However, the size of the $1\sigma$ error bar is large enough to give a good agreement between the best-fitting and fiducial values, especially for the range $s_{\rm min} = 40\Mpch$. 
From the $Q(s)$ panels of Fig.~\ref{fig:beta_lin_s1}, we observe an underestimation of $\beta$ for all gravity models at all measured redshifts, especially for the range $s = 20 - 150 \Mpch$. As we said above, non-linearities produce smaller values of $Q(s)$ at all redshifts and therefore we estimate a smaller value of $\beta$ even on scales with $s_{\rm min} = 40\Mpch$. When we use clustering wedges, $\xi_w(s)$, to estimate $\beta$ (see the bottom panels of Fig.~\ref{fig:beta_lin_s1}), we find a similar trend consistent to that seen for $Q(s)$. This is because there is a relation between clustering wedges and multipole moments (see Eq.~\ref{eq:xiw_l}). When we measure $Q(s)$, we only use information about the monopole and the quadrupole moments of the correlation function, while the linear prediction of the clustering wedges uses information about the monopole, quadrupole and hexadecapole moments. 
The comparison between constraints using $\xi_w(s)$ and $Q(s)$ therefore indicates that the hexadecapole moment does not have much impact on the estimation of $\beta$. In general, the linear Kaiser model fails to model RSD in configuration space even in the linear regime $(s_{\rm min} = 40\Mpch)$.

The lower subpanels of each plot in Fig.~\ref{fig:beta_lin_s1} show the relative difference between the MG models and GR. We notice that in all cases the difference between $f(R)$ models (F6 and F5) is $\lesssim 1\%$ with respect to GR, making these models statistically indistinguishable from each other. On the other hand, N5 and N1 models hold a difference of $\sim 2.5\%$ and $\sim 12\%$ with respect to GR. Also, while the $R(s)$, $Q(s)$ and $\xi_w(s)$ estimators all lead to biased constraints on $\beta$ for all models and redshifts, it appears that the bias is the same for the different gravity models such that the relative model differences from GR can be more accurately recovered.

In Appendix~\ref{append:cov_mat} we will show the estimation of $\beta$ by using an alternative method to estimate the error budget in the $\chi^2$-test.

\subsection{The nonlinear model}\label{sec:gRPT}

A more rigorous and accurate modelling of the clustering signal of galaxies in redshift space than the linear theory prediction can be achieved by accounting for three important ingredients: the nonlinear evolution of the underlying matter field, the redshift space distortion effects, and the biasing relation between the galaxy and dark matter fields. Within the redshift range we are concerning in this work, apart from the linear theory prescription described above, one needs in principle to include also the cross coupling between the matter field with the velocity field, the higher order and nonlocal bias to account for the nonlinear and nonlocal formation process as well as a modelling of the virialized random motion of the objects (galaxies).
The anisotropic correlation function can be obtained as the Fourier transform of the power spectrum, where the full expression of $P(k,\mu)$ is given by
\begin{equation}\label{eq:Pk_grpt}
P(k,\mu) = F_{\rm FoG}(k,\mu)P_{\rm novir}(k,\mu)\,,
\end{equation}
where
\begin{equation}\label{eq:F_fog}
F_{\rm FoG}(k,\mu) = \frac{1}{\sqrt{1 + f^2k^2\mu^2a^2_{\rm vir}}}\exp{\left(\frac{-f^2k^2\mu^2\sigma^2_v}{1 + f^2k^2\mu^2a^2_{\rm vir}}\right)}\,,
\end{equation}
is a non-Gaussian term that contains small-scale information about the Fingers-of-God effect due to virialised motions of galaxies, $a_{\rm vir}$ is a free parameter that describes the {kurtosis of the velocity distribution on small scales}  and $\sigma_v$ is the velocity dispersion \citep{Scoccimarro:2004tg}.
The non-virial power spectrum $P_{\rm nonvir}$ is given by three contributions,
\begin{equation}\label{eq:P_novir}
P_{\rm novir}(k,\mu) = P^{(1)}_{\rm novir}(k,\mu) + (k\mu f)P^{(2)}_{\rm novir}(k,\mu) + (k\mu f)^2P^{(3)}_{\rm novir}(k,\mu),
\end{equation}
where
\begin{eqnarray}
P^{(1)}_{\rm novir}(k,\mu) &=& P_{\rm gg}(k) + 2 f \mu^2 P_{\rm g \theta}(k) + f^2\mu^4 P_{\theta \theta}(k)\,,\label{eq:P1_rpt}\\
P^{(2)}_{\rm novir}(k,\mu) &=& \int \frac{{\rm d}^3 p}{(2\pi)^3}\frac{p_z}{p} [B_\sigma({\bf p},{\bf k - p}, {\bf -k}) - B_\sigma({\bf p},{\bf k}, {\bf -k - p})]\,,\nonumber \\
&~&\label{eq:P2_rpt}\\
P^{(3)}_{\rm novir}(k,\mu) &=& \int \frac{{\rm d}^3 p}{(2\pi)^3} F({\bf p}) F({\bf k - p}) \,,\label{eq:P3_rpt}
\end{eqnarray}
Here, $P^{(1)}_{\rm novir}$ is a non-linear version of the Kaiser formula, Eq.~\eqref{eq:Pk_s_lin}, {$P_{\rm gg} \equiv \left\langle \delta_{\rm g} \delta_{\rm g}\right\rangle$ is the galaxy auto power spectrum, $P_{\rm g \theta} \equiv \left\langle \delta_{\rm g}{\theta}\right\rangle$ is the cross spectrum between the galaxy density ($\delta_{\rm g}$) and velocity divergence ($\theta\equiv\nabla\cdot{\bf v}$, and assuming there is no velocity bias) fields,and $P_{\theta\theta} \equiv \left\langle {\theta}{\theta}\right\rangle$. The exact expressions for the first two terms can be found in Appendix A of \citealt{Sanchez:2016sas}. The calculation of the nonlinear matter power spectrum is done using the Galilean-invariant renormalized perturbation theory (gRPT; \citealt{Crocce:2005xy}; Crocce, Blas \& Scoccimarro in prep.)} { 
When calculating the ensemble average of the product of the density field in redshift space (c.f. Eq.~(15) of \citet{Taruya:2010BAO}), there is coupling between the FoG and the Kaiser effect as $P^{(2)}_{\rm novir}$ and $P^{(3)}_{\rm novir}$ and add correction to the Kaiser term.} $P^{(2)}_{\rm novir}$ is given by the tree-level PT bispectrum between densities and velocities. $P^{(3)}_{\rm novir}$ is the quartic linear power spectrum { at the order $\mathcal{O}\left( P^2\right)$ with the kernel $F({\bf{P}})=\frac{p_z}{p^2}\left(P_{\delta\theta}(p)+ f\frac{p^2_z}{p^2} P_{\theta\theta}(p)\right)$}. Eq.~(\ref{eq:Pk_grpt}) encodes effects of both the RSD and nonlinear evolution. 

The galaxy bias in this model is expanded as follows \citep{Chan:2012jj},
\begin{equation}\label{eq:b_rpt}
\delta_{\rm g} = b_1 \delta + \frac{b_2}{2}\delta^2 + \gamma_2 \mathcal{G}_2 + \gamma^{-}_3 \Delta_3 \mathcal{G} + ...
\end{equation}
with the Galileon operators for the gravitational potential $\Phi$ and the velocity potential $\Phi_{\rm v}$ defined as
\begin{equation}
\mathcal{G}_2\left(\Phi_v\right) \equiv \nabla^i\nabla^j\Phi_v\nabla_{i}\nabla_{j}\Phi_v - \left(\nabla^2\Phi_v\right)^2,
\end{equation}
and
\begin{equation}
\Delta_3 \mathcal{G} \equiv \mathcal{G}_2\left(\Phi\right) - \mathcal{G}_2\left(\Phi_v\right).
\end{equation}
The non-local bias coefficients $\gamma_2$ and $\gamma^-_3$ are related to the linear bias parameter, $b_1$, as \citep{Fry:1996fg,Catelan:1997qw,Chan:2012jj}
\begin{equation}
\gamma_2 = -\frac{2}{7}(b_1 - 1),
\end{equation}
and
\begin{equation}\label{eq:g3_ll}
\gamma^-_3 = \frac{11}{42}(b_1 - 1).
\end{equation}
We have tried both fixing and varying $\gamma^-_3$ in our fitting. When fixing thie parameter using the local Lagrangian relation, Eq.~\eqref{eq:g3_ll}, we found the linear bias is biased low compared to the true value. This behaviour could due to the fact that $\gamma^-_3$ is formulated in the Eulerian coordinate, while the local biasing schemes are compatible with the Lagrangian bias only when matter evolution and structure formation is well within the linear and local regime \citep{Matsubara:2011}. 
Therefore varying $\gamma^-_3$ should results in a more accurate value and this conclusion is consistent with the previous findings \citep{Grieb:2017dr12}. In the results below we shall always vary $\gamma_3^-$.

\subsubsection{The Alcock-Paczynski effect}\label{sec:AP}

The baryon acoustic oscillation can be well approximated by a spherical shape with fixed radius at given redshift. As one measures the clustering signal parallel and perpendicular to the line of sight, a set of parameters known as the Alcock-Paczynski (AP; \citealt{Alcock:1979mp}) parameters can be introduced to account for the rescaling of the BAO feature in both the radial and angular directions:
\begin{equation}
q_\perp = \frac{D_A(z_{\rm m})}{D_A'(z_{\rm m})}\,, \qquad
q_\parallel = \frac{H'(z_{\rm m})}{H(z_{\rm m})}\,,
\end{equation}
where the $\prime$ denotes quantities in the fiducial cosmology. 
In terms of $s$ and $\mu$, these equations can be written as
\begin{equation}
s = s^{\prime}q(\mu^{\prime})\,, \qquad
\mu = \mu^{\prime}\frac{q_{\perp}}{q(\mu^{\prime})}\,,
\end{equation}
where 
\begin{equation}
q(\mu') = \sqrt{q_{\parallel}^2 (\mu^{\prime})^2 + q_{\perp}^2 \left( 1 - (\mu^{\prime})^2\right)}.
\label{eqn: AP_s_mu}
\end{equation}
With Eq.~(\ref{eqn: AP_s_mu}), the correlation function predicted by the model for a fiducial cosmology can be transformed into the prediction for different cosmologies $\xi(s',\mu') \rightarrow \xi(s,\mu)$.
When measuring the two point correlation function, we have used the true position of the objects. Since the expansion history is tuned to be the same for each cosmological model, we effectively always know the "true cosmology", and would therefore expect to find $q_\perp=q_\parallel=1$ for all the cases.

As we will see in Appendix~\ref{append:nuisance_params}, the constraints on AP parameters for different cosmological models are very close to one, this is a good news when applying to a real survey. Despite the shape of the distortion at all range of scales due to the RSD, the AP test can faithfully pick up the correct information given by the BAO position.

\begin{figure*}
 \centering
\includegraphics[width=0.45\textwidth]{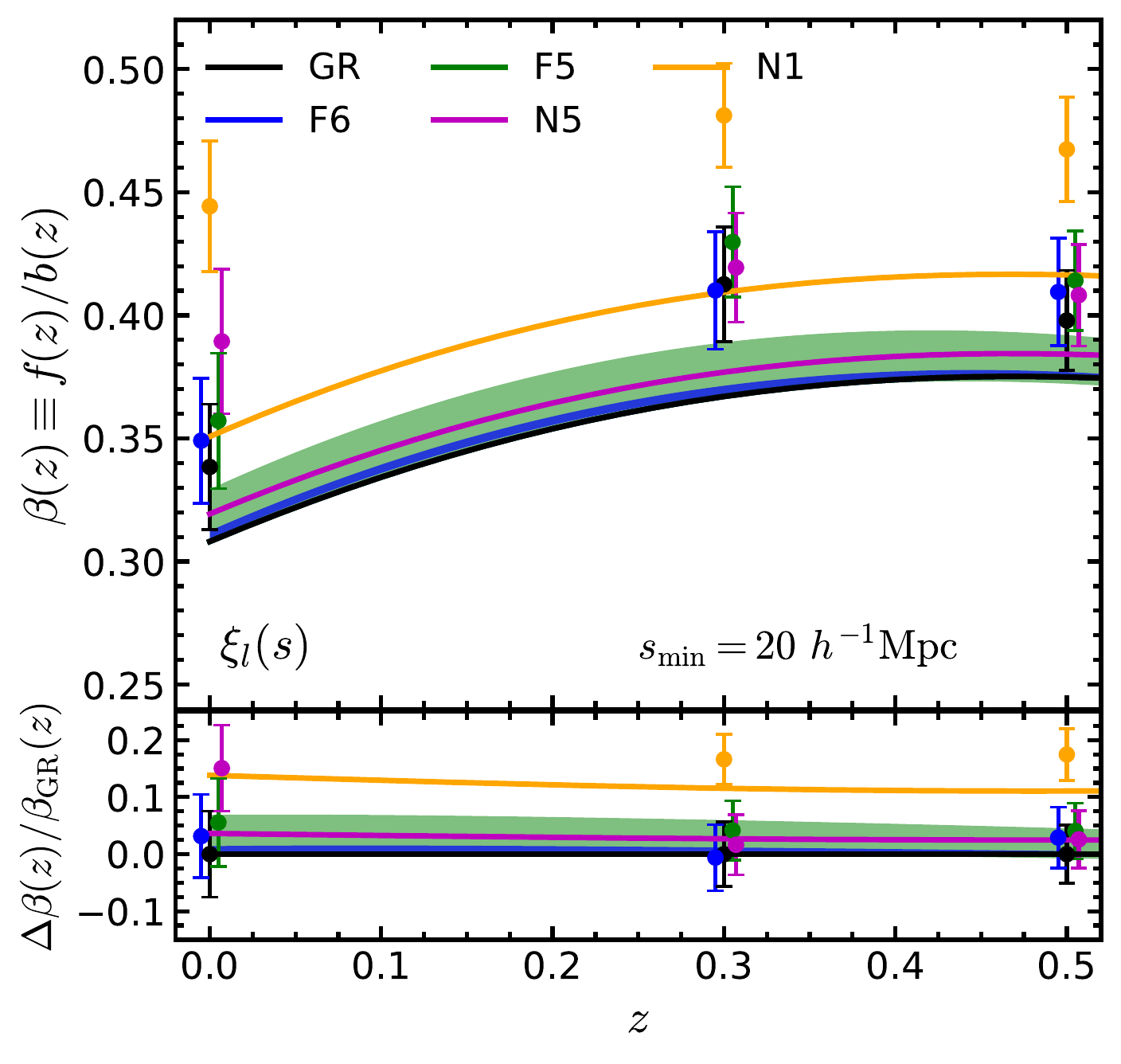}
\includegraphics[width=0.45\textwidth]{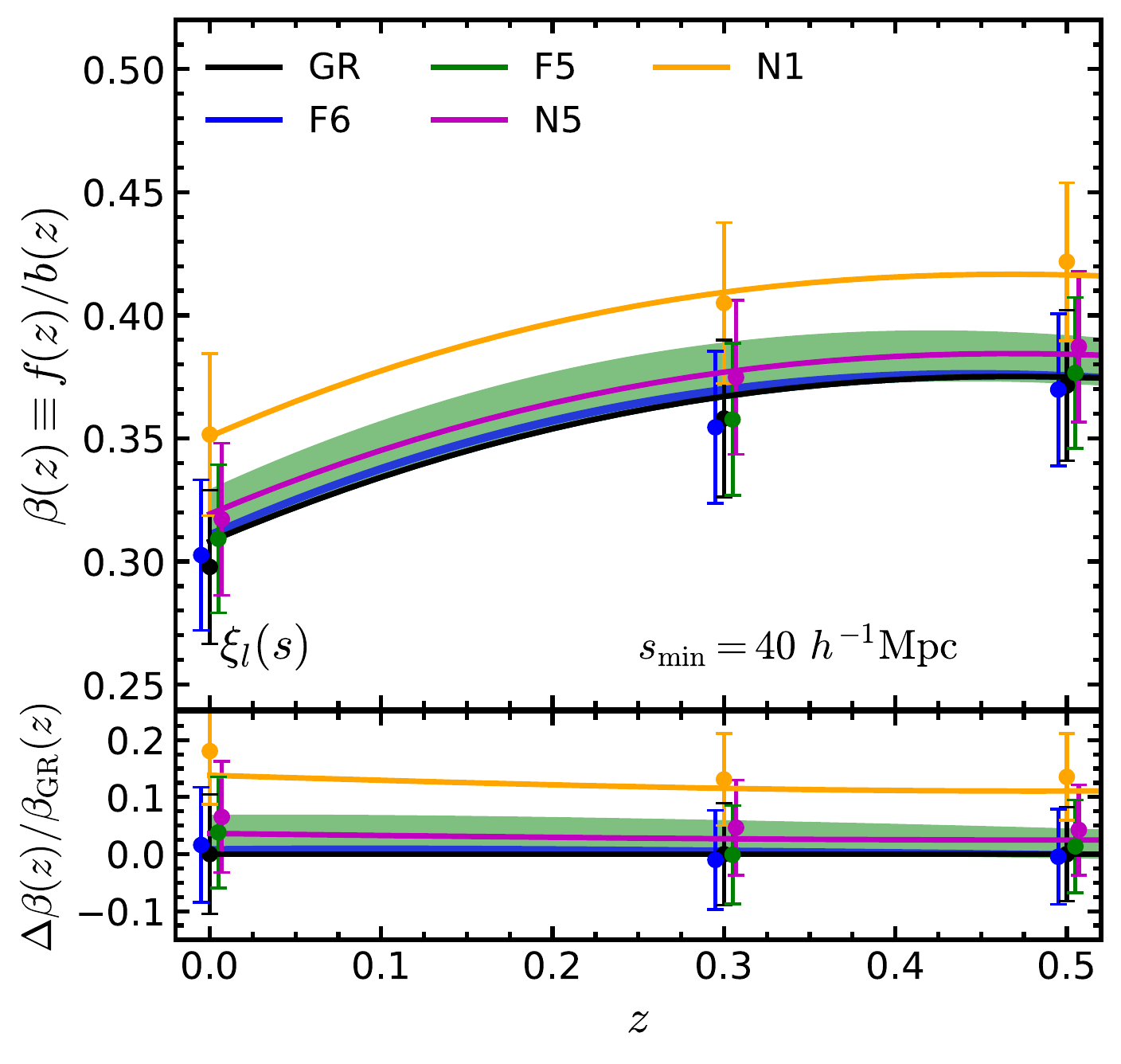}
\includegraphics[width=0.45\textwidth]{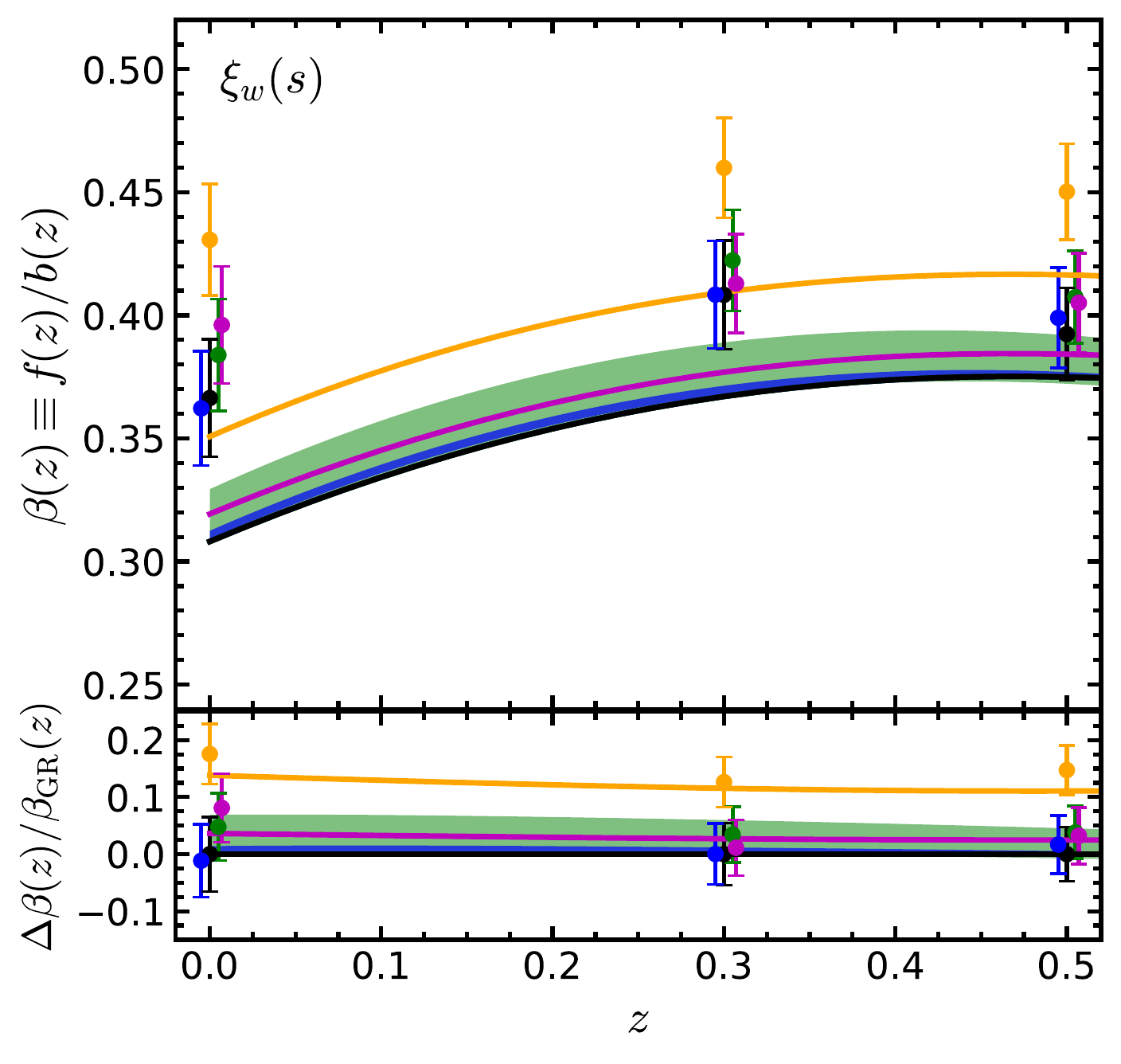}
\includegraphics[width=0.45\textwidth]{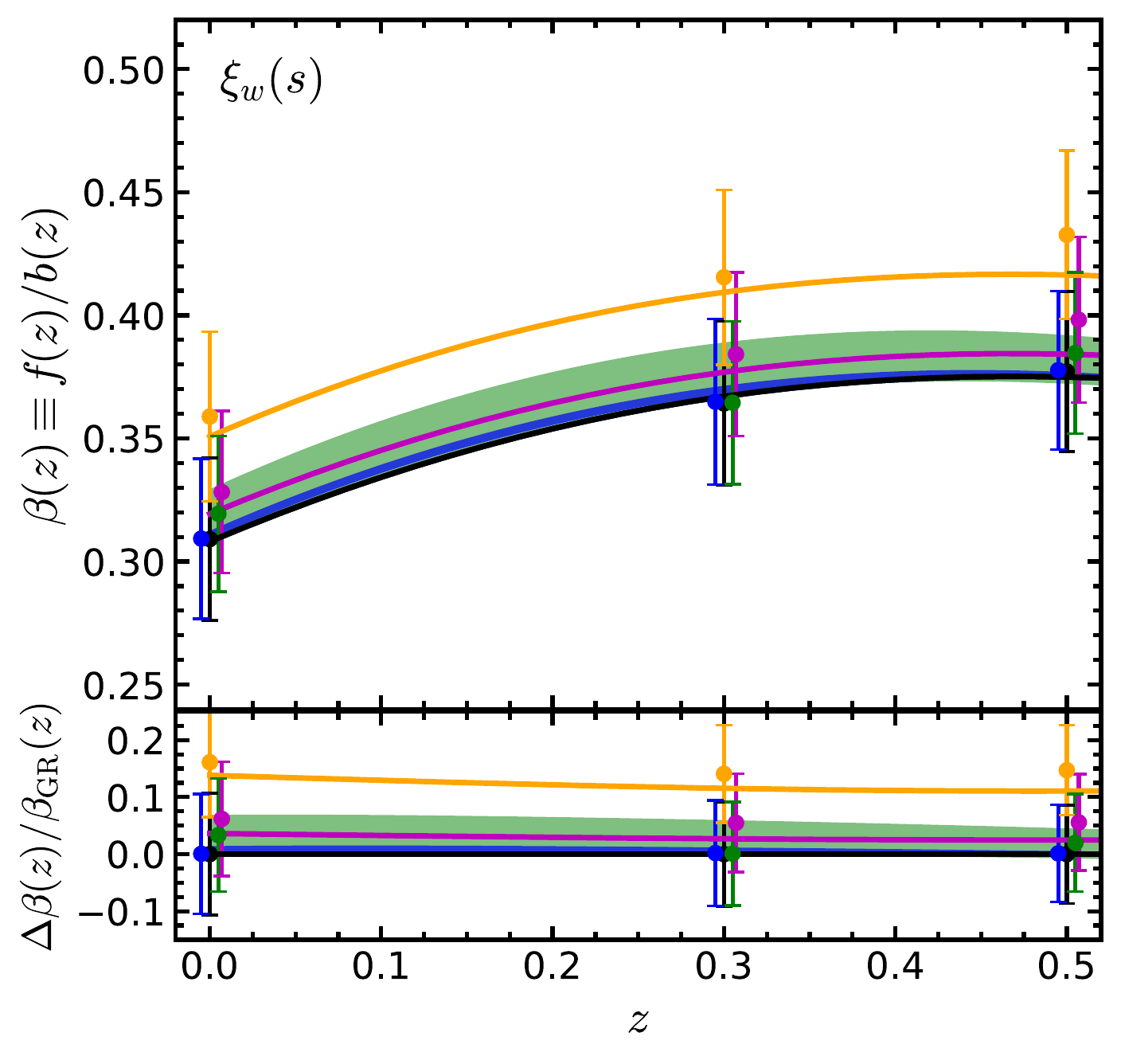}
	\caption{Similar to figure~\ref{fig:beta_lin_s1} but for the fits using the nonlinear model. The upper panel shows the multipole moments of the correlation functions, $\xi_l(s)$. The lower panel shows the clustering wedges $\xi_{w_i}(s)$. In both of the plots the AP parameters are treated as free.}
	\label{fig:beta_freeAP}
\end{figure*}

\subsubsection{Parameter inference with the nonlinear model}\label{sec:params inference}
To obtain cosmological constraints, we use Bayesian statistics and maximise the likelihood,
\begin{equation}
\mathcal{L}({\bm \xi}|{\bm \lambda})\propto \exp\left[-\frac{1}{2}\left({\bm \xi}-{\bm \xi}_{\rm model}({\bm \lambda})\right)^{\rm T}{\Psi}\left({\bm \xi}-{\bm \xi}_{\rm model}({\bm \lambda})\right)\right],
\label{eqn: gaussian-likelihood}
\end{equation}
where the $\Psi = \rm C^{-1}$ is the inverse of the covariance matrix. We applied the Gaussian recipe to estimate the covariance matrix \citep{Grieb:2016cov}, which is then rescaled by the number of simulations. The input power spectrum is calculated by the nonlinear model based on the best fitting values obtained from the MCMC chain. Such Gaussian recipe has been tested recently in both Fourier and configuration space by comparing to covariance matrices generated by hundreds of N-body simulations as well as thousands of different fast mock simulations and found them to be in good agreement \citep{Blot:2018cov, Lippich:2018cov}. At the same time, there are also studies on including the corrections from higher-order statistics and super-sampling mode \citep{Barreira:2018cov}. 
However, for the scales of interest in this study, there is no sensitive to these correction and the Gaussian covariance matrix should be a good approximation.
We explore the parameter space using Monte Carlo Markovian Chains (MCMC) with the Metropolis-Hastings algorithm \citep{Metropolis:1953,Hastings:1970}. The parameters that enter the default fitting are $\{f\sigma_8, b_1, b_2, \gamma^-_3, a_{\rm vir}\}$. When applying the AP test, two additional parameters enter the fitting $\{\alpha_\parallel, \alpha_\perp\}$. Finally, we marginalise over the nuisance parameters to find the probability distribution of the distortion parameter $\beta = f/b_1$.

Fig.~\ref{fig:beta_freeAP} shows the constraints on $\beta$ using the nonlinear gPRT+RSD model by running MCMC. The upper panels present the results for the three multipoles ($\xi_l(s)$, $l=0,2,4$) of the correlation function for two ranges of scales: $s=20-150\Mpch$ (left) and $s=40-150\Mpch$ (right). 
For comparison, we display the results of using three wedges ($\xi_w(s)$) in the bottom panels of Fig.~\ref{fig:beta_freeAP}. We observe an overestimation of $\beta$ for all models at all redshifts when the fit is done using $s_{\rm min} = 20\Mpch$, for both multipoles and wedges. We have checked the linear bias fitted from the nonlinear model and found it to be in good agreement with the values measured from the mock galaxy catalogues using Eq.~(\ref{eq:bias}); for a detailed discussion on the bias see Appendix~\ref{append:bias}. 
This suggests that the higher estimation of $\beta$ comes from the $f\sigma_8$; the same conclusion is in agreement with the one found by \citet{Barreira:2016ovx}. In our case we have rescaled the covariance matrix by the number of simulations and the error bar is therefore smaller than the error bar presented in \citet{Barreira:2016ovx}. When we used the range scale with $s_{\rm min} = 40\Mpch$, the constraints are in good agreement with the fiducial values. 

We note a slight difference between the results obtained from the multipoles-based estimator and the wedges-based estimator. This is an indication for the non-negligible higher order statistics in the two-dimensional correlation function $\xi(s,\mu)$. To further explore this point, we have compared the difference between the multipoles {directly measured from the mock catalogues and the multipoles obtained by transforming the measured wedges using Eq.~(\ref{eq:wedge-transform}), and found a difference in the hexadecapole at scale $\approx 20\Mpch$ (we do not show the plot here). The difference can lead to different constraints on the nuisance parameters and have consequential impact on the parameter of interest, which is a source for the biased constraints by setting $s_{\rm min} = 20 \Mpch$. For the case of $s_{\rm min}=40\Mpch$, the minimum length scale used in the fitting is larger enough, and the impact of higher-order statistics is smaller, which explains why the agreement is improved at all three redshifts.

In the lower subpanels of Fig.~\ref{fig:beta_freeAP} we plot the relative differences between the modified gravity models and GR. Similar to the findings using linear theory, we find that the two variants of $f(R)$ gravity studied here are indistinguishable from GR given the size of the statistical error. While N5 shows stronger deviation from GR, it is also not clearly distinguishable from the latter. N1 is the only one of our four MG models that could be distinct from GR at 1-$\sigma$ given the statistical uncertainties, and not including systematic errors. These results are similar to what we found by using the linear Kaiser model above. 

\subsection{Discussion}
\label{sec:result_discuss}
The results can have a few implications:

First, RSD on linear and mildly-nonlinear scales does not seem to be a great probe of modified gravity, in particular for $f(R)$ gravity. This conclusion is expected to hold true for other MG models depending on chameleon screening to recover GR in high-density regions, for which the effect of the fifth force is generally restricted to at most $\mathcal{O}(10)$ Mpc \citep{Brax:2012gr,Brax:2011aw}. This conclusion, however, may not apply to RSD on small and highly nonlinear scales, where the velocity field could be significantly enhanced by the fifth force in chameleon models \citep[see, e.g.,][]{He:2018oai}. We suspect similar conclusions should hold for the symmetron \citep{Hinterbichler:2010es} and dilaton \citep{Brax:2010gi} models, for which the fifth force is also of the Yukawa type, with an inverse Compton mass of $\lesssim\mathcal{O}(10)$ Mpc. Vainshtein screening models, such as nDGP, on the other hand, has a fifth force that is non-negligible on large scales, which is why the constraint is stronger.

Second, given the weak constraining power from large-scale RSD and the relatively large scale ($s_{\rm min}=40\Mpch$) needed to get unbiased constraints even for GR, a theoretical model based on linear theory prediction or higher-order perturbation calculation developed for GR does not seem to lead to noticeably biased constraints on the $\beta$ parameter. This suggests a faster way to explore the MG model and parameter space, at least at the initial stage of delineating models and parameters.

Third, we have explicitly checked that the real-space galaxy correlation functions of the MG models deviate more significantly from GR prediction if the mock galaxy catalogues were constructed using the same HOD parameters as GR, or if haloes were used instead of HOD galaxies. As argued above, in this study the HOD parameters for MG models are tuned so that the real-space $\xi_{\rm gg}$ match between the different models, which is motivated by the fact that there is only one Universe from which the observed $\xi_{\rm gg}$ are to be derived, and whatever cosmological model should be required to reproduce such observation to start with. A more detailed theoretical model of RSD on linear and mildly-nonlinear scales should take this into account. In practice, there is no real-space $\xi_{\rm gg}$ from observations to match with, and the HOD parameters are often tuned to match the observed projected two-point correlation function $w(\theta)$ to get rid of RSD effects: doing this will leave more freedom for the choices of HOD parameters, and we expect it to also lead to larger uncertainties in the constraints on $\beta$.

\section{Summary and conclusions}\label{sec:conc}
We have presented results on the estimation of the redshift space distortion parameter, $\beta$, which is directly related to the linear growth rate of matter in the Universe. In order to understand the potential of using this parameter to constrain cosmological models, we have tested the estimation of this parameter for five different gravity models: a flat $\Lambda$CDM model based on General Relativity (GR), two variants of the \citet{Hu:2007nk} $f(R)$ gravity model (F6 and F5) and two variants of the normal branch of the DGP \citep{Dvali:2000hr} model (N5 and N1). The objective of this study is threefold: first, we want to explore whether for realistic mock galaxy catalogues the $\beta$ parameter is sufficiently different between the different gravity models so that future galaxy surveys can be used to distinguish or constrain them; second, we study the extent to which simple theoretical models such as linear theory or GR-based perturbation theory recipes can faithfully recover the correct $\beta$ values for different MG models, given the current statistical uncertainties; finally,  we also compare different estimators of the RSD effect and test various systematic effects in modelling RSD.

To do so, we use cosmological dark-matter-only N-body simulations and populated dark matter haloes with galaxies following a halo occupation distribution prescription. We did this analysis for three low redshifts, respectively at $z=0$, $0.3$ and $0.5$, because the modified gravity models studied here are expected to deviate from GR more significantly at late times. Since the nature of gravity is different in every model, we tuned the HOD parameters such that essentially every catalogue matches the number density and the real space correlation function measured for the {\sc boss cmass dr9} \citep{Anderson:2012sa,Manera:2012sc}. We used the distant-observer approximation to map galaxies from real- to redshift-space coordinates along three line-of-sight directions (chosen to be parallel to the three axes of the simulation box) for each realisation of mock galaxy catalogue. 
For the theoretical predictions of the RSD effects, we applied a linear \citep{Kaiser:1987,Hamilton:1992zz} and a nonlinear (\citealt{Scoccimarro:2004tg,Crocce:2005xy,Chan:2012jj}; Crocce et al., in prep.) RSD model to our mocks to estimate the value of $\beta$. We used different estimators to extract information about the distortion parameter in each model. In linear theory we have the ratios $R(s)$ and $Q(s)$ besides the clustering wedges $\xi_w(s)$. For the nonlinear model we have used the multipole moments $\xi_l(s)$ and the clustering wedges of the correlation function. For both RSD models we performed fits over two ranges of scales, $s=20-150\Mpch$ and $s=40-150\Mpch$.

In general, we found that the linear model fails to recover the true value of the distortion parameter even in the linear regime of scales ($s_{\rm min} = 40\Mpch$). This is because, due to the lack of the FoG term in this model, it  (over-)underestimates the value of $\beta$ on the quasi-linear regime ($s_{\rm min} = 20\Mpch$). On the other hand, the nonlinear model overestimates the value of $\beta$ when considering  $s_{\rm min} = 20\Mpch$; this trend was also found by \citet{Barreira:2016ovx} using the same range scale. In the linear regime ($s_{\rm min} = 40\Mpch$), the nonlinear model is able to recover the true value of $\beta$, especially for clustering wedges. 

Our main conclusions are as follows:
\begin{enumerate}
\item Measurements of redshift space distortions on large scales can help us to distinguish between some gravity models, such as N1, but in general the model differences from GR are small compared with statistical and theoretical uncertainties, in particular for $f(R)$ model or chameleon models in general.
\item Chameleon and Vainshtein models have distinct model predictions, which are directly related to the different properties of the fifth forces in the models: in chameleon-type models the fifth force is of Yukawa type and gets suppressed on scales above the inverse Compton mass of the scalar field (typically $\sim10$ Mpc or smaller), while for Vainshtein-type models the fifth force is long ranged and can alter the large-scale velocity field substantially.
\item The lack of a Fingers-of-God term in the linear Kaiser model produces an over and underestimation of the $\beta$ parameter when used $R(s)$ and $Q(s)/\xi_w(s)$, respectively.
\item The linear Kaiser prediction is independent of the model of gravity, while the nonlinear model in its current form is based on GR only and we have tested it in $f(R)$ gravity models for the first time (the same nonlinear model was used to validate estimations of the growth rate for nDGP models in \citet{Barreira:2016ovx}). The fact that the use of this `incorrect' non-linear model produces reasonable constraints for $\beta$ for the modified gravity models studied here offers a practical way to measure possible signatures of modified gravity in the large-scale structure of the Universe. 
\item We have tested the effect of using different ranges of scales in the fitting, and found that for $s_{\rm min}=20\Mpch$ the nonlinear model cannot recover $\beta$ correctly at all redshifts for all models including GR (a result similar to the findings of \citet{Barreira:2016ovx}; note that the relative model differences from GR can be better recovered in this case), but the issue can be resolved by excluding data between $20$ and $40\Mpch$, at the expense of increased uncertainties.
\item Using different estimators such as multipoles and wedges can produce different constraints because of the different information they encode, but the difference is not statistically significant.
\end{enumerate}

We note that our conclusions are different from other recent works, such as \citet{He:2018oai}. This is due to the focus on different scales (\citeauthor{He:2018oai} concentrated on small and highly nonlinear scales), and reflects the strong scale-dependence of the behaviours in some MG models.

Our results suggest that, with the upcoming galaxy surveys such as {\sc desi}, {\sc 4most} and {\sc Euclid}, there is a realistic possibility to put constraints on the growth rate of matter and make distinctions between certain MG models and GR. Such studies will potentially benefit from combining with cosmological data that probe different regimes (e.g., environments), scales and special theoretical properties of the models. Also, to be more realistic, it will be useful to test the constraining power of RSD using different tracers and number densities, and include systematic effects such as survey geometric and masks, galaxy redshift distribution and evolution, incompleteness due to fibre collisions and observing conditions, and so on. It is also interesting to study if including higher-order statistics, such as the 3-point correlation function or bispectrum in redshift space, can improve the constraining power of the surveys. These possibilities are beyond the scope of this paper and will be left for future work.

\section*{Acknowledgements}

The authors wish to thank Roman Scoccimarro and Martin Crocce for developing the nonlinear model used in this work.
CH-A acknowledges support from the Mexican National Council of Science and Technology (CONACyT) through grant No. 286513/438352. 
JH acknowledges useful discussion with Alexandre Barreira, Daniel Farrow, Martha Lippich and Fabian Schmidt.
BL is supported by the European Research Council (ERC-StG-716532-PUNCA). 
We acknowledge support from STFC Consolidated Grants ST/P000541/1, ST/L00075X/1. 
This work used the DiRAC Data Centric system at Durham University, operated by the Institute for Computational Cosmology on behalf of the STFC DiRAC HPC Facility (\url{www.dirac.ac.uk}). This equipment was funded by BIS National E-infrastructure capital grant ST/K00042X/1, STFC capital grants ST/H008519/1 and ST/K00087X/1, STFC DiRAC Operations grant ST/K003267/1 and Durham University. DiRAC is part of the National E-Infrastructure.



\bibliographystyle{mnras}
\bibliography{ref} 



\appendix
\section{Tests of systematic effects}

\begin{figure*}
 \centering
\includegraphics[width=0.45\textwidth]{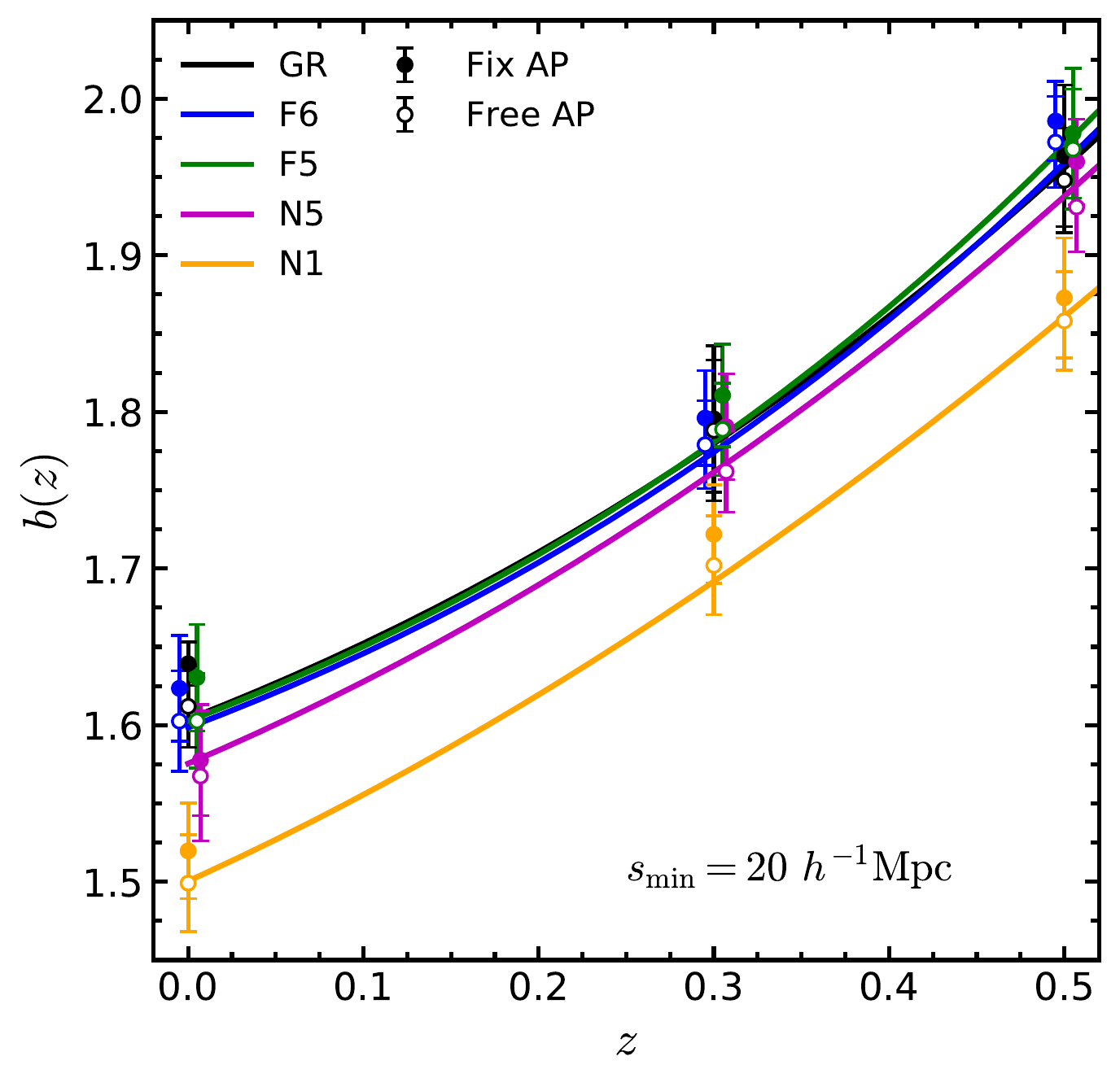}
\includegraphics[width=0.45\textwidth]{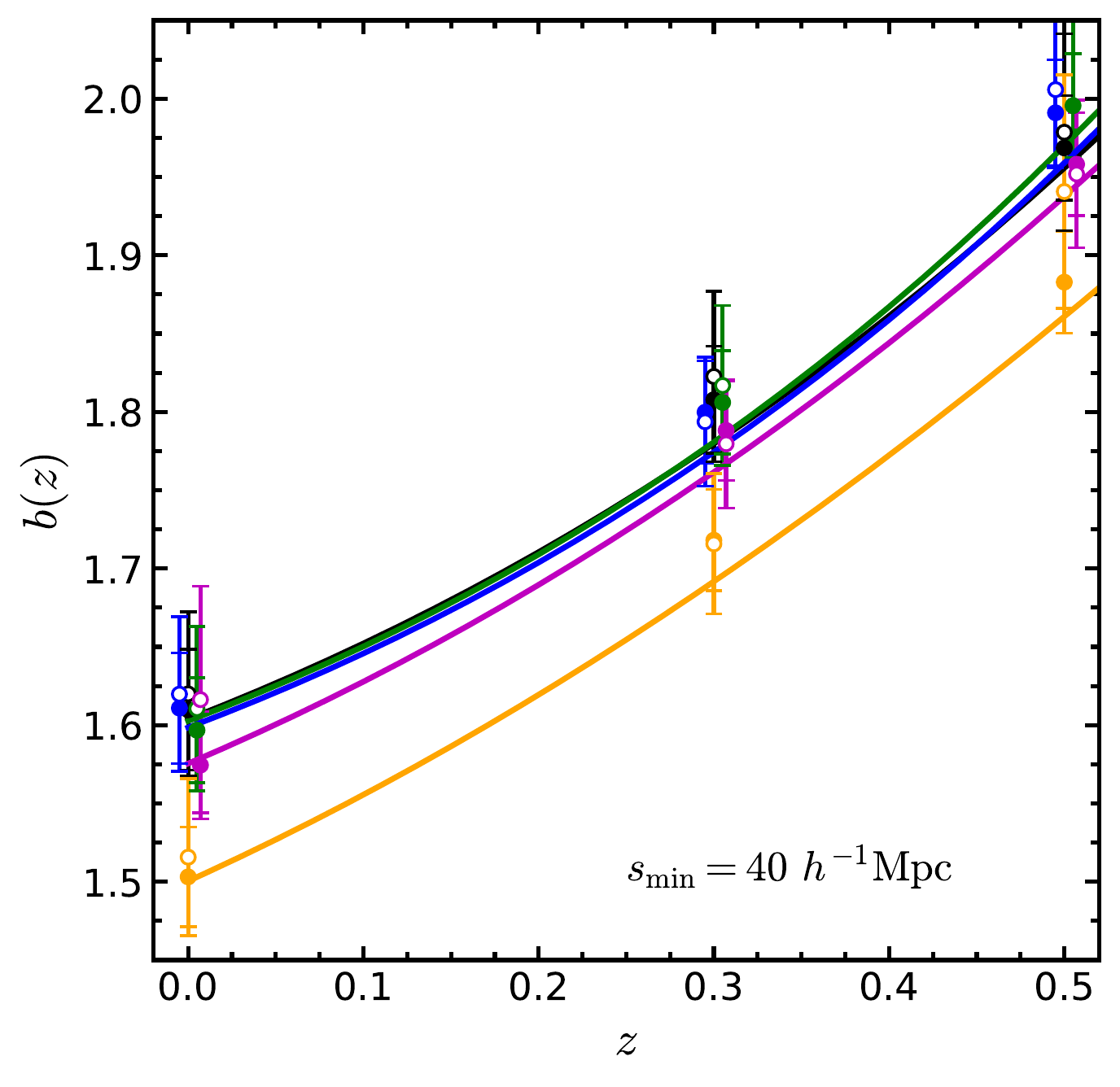}
	\caption{A comparison of the linear bias parameter $b_1$ obtained by appropriately rescaling the best-fit value $b_{1_{\rm MCMC}}$ using the nonlinear model (symbols) and from direct measurements using the mock galaxy catalogues (lines), for the three redshifts (as shown by the horizontal axis) and all models (see legend). The filled and open symbols are respectively from MCMC fittings where the AP parameters $q_\perp,q_\parallel$ are fixed and are left free to vary; the left and right panels are respectively from MCMC fittings with $s_{\rm min}=20$ and $40h^{-1}$Mpc.}
	\label{fig:bias_gRPT}
\end{figure*}

In this Appendix we include some tests of the systematic effects in our constraints, and materials that contain additional information to the results shown in the main text.

\subsection{Systemtics validation: linear bias}
\label{append:bias}

Figure \ref{fig:bias_gRPT} shows the linear bias parameter derived from the fitting using the nonlinear model compared to the actual values measured from the mock galaxy catalogues. The left panel shows the bias values obtained by using a minimum fitting scale $s_{\rm min} = 20\Mpch$ and the right panel corresponds to using $s_{\rm min} = 40\Mpch$. In the MCMC fitting, the matter power spectrum was calculated by calling the {\sc camb} code with an input of fiducial $\sigma_8$. The amplitude of the matter power spectrum is therefore degenerate with the linear bias the $b_1$. In Figure \ref{fig:bias_gRPT}, the linear bias is obtained by a rescaling, $b_1 = b_{1_{\rm MCMC}}\sigma_8^{\rm input}/\sigma_8^{\rm true}$. The initial conditions of our simulations were generate using Zel'dovich approximation at $z_{\rm ini}$, which has a worse-than-percent-level accuracy \citep{Crocce:2006ve}; therefore the $\sigma_8^{\rm true}$ value we used in this rescaling is $0.844$, which was obtained by requiring that the resulting {\sc camb} power spectrum to best agree with the one measured directly from the N-body initial condition.

The filled and open symbols in Figure \ref{fig:bias_gRPT} are respectively the linear bias $b_1$ for the different models and redshifts rescaled using the corresponding the constraints on $b_{1_{\rm MCMC}}$ values where the AP parameters are fixed and left free to vary during the MCMC fitting. We can see that in both cases they agree well with the true results measured from the mock galaxy catalogues (the coloured curves) for GR, F6, F5 and N5. For N1 the constraint on the bias values are significantly higher compared with the true values, which is because the $\sigma_8^{\rm true}$ used in the rescaling is the GR value, and the corresponding N1 value is larger (for the other MG models the difference of $\sigma_8^{\rm true}$ from GR is smaller). We also find that the $b_1$ values are well recovered for both $s_{\rm min}=20$ and $40\Mpch$.

\begin{figure*}
 \centering
\includegraphics[width=0.33\textwidth]{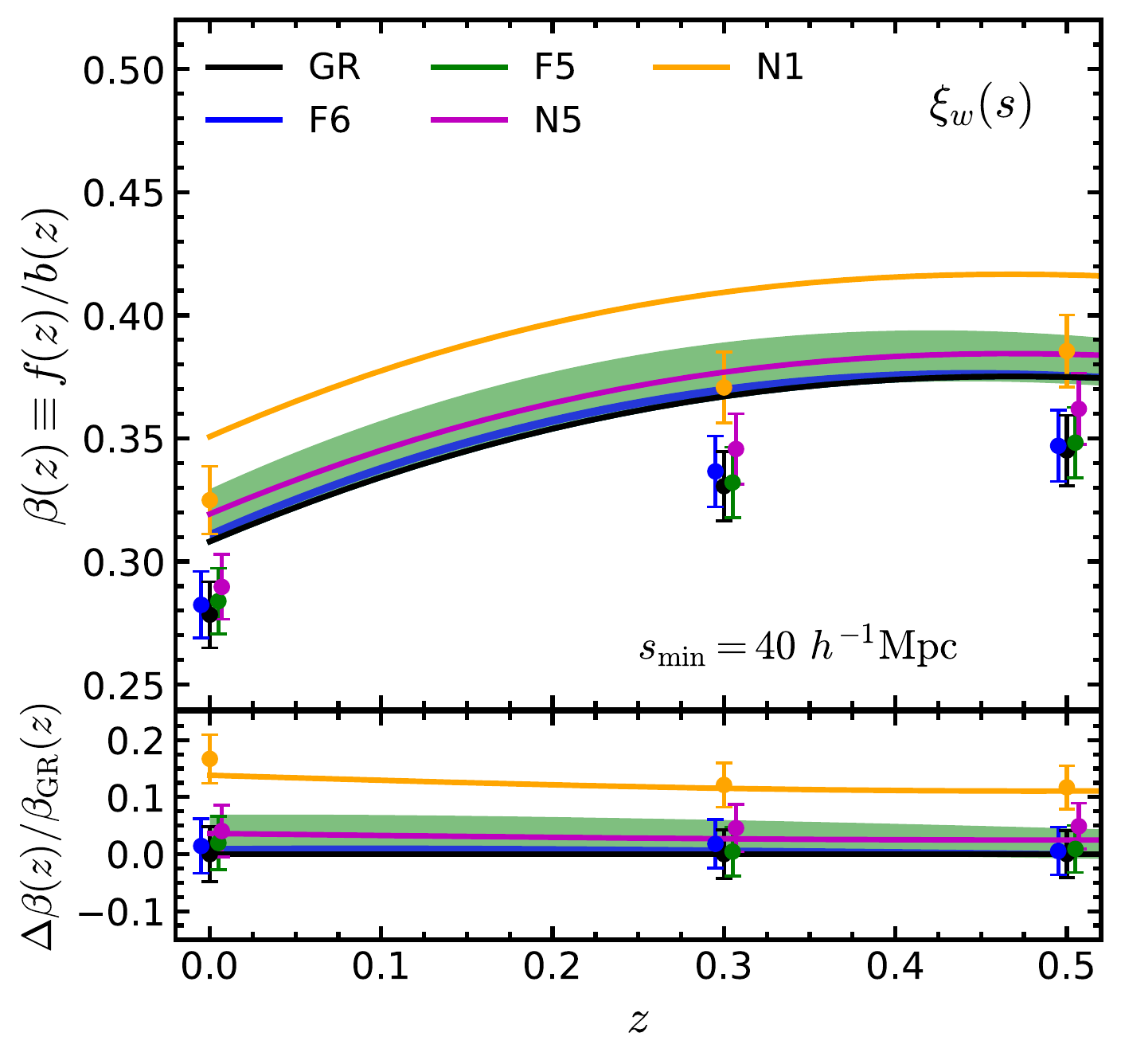}
\includegraphics[width=0.33\textwidth]{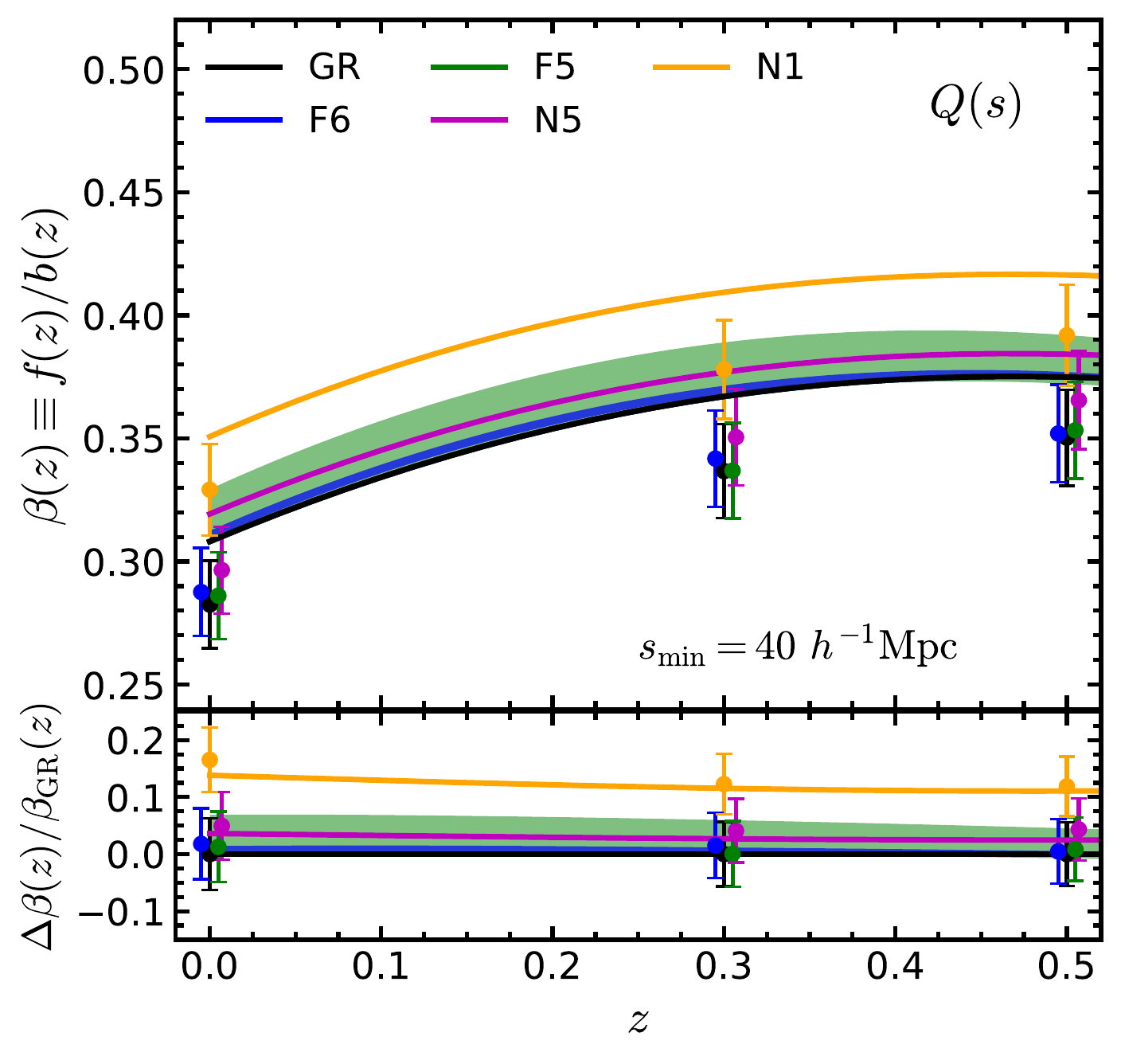}
\includegraphics[width=0.33\textwidth]{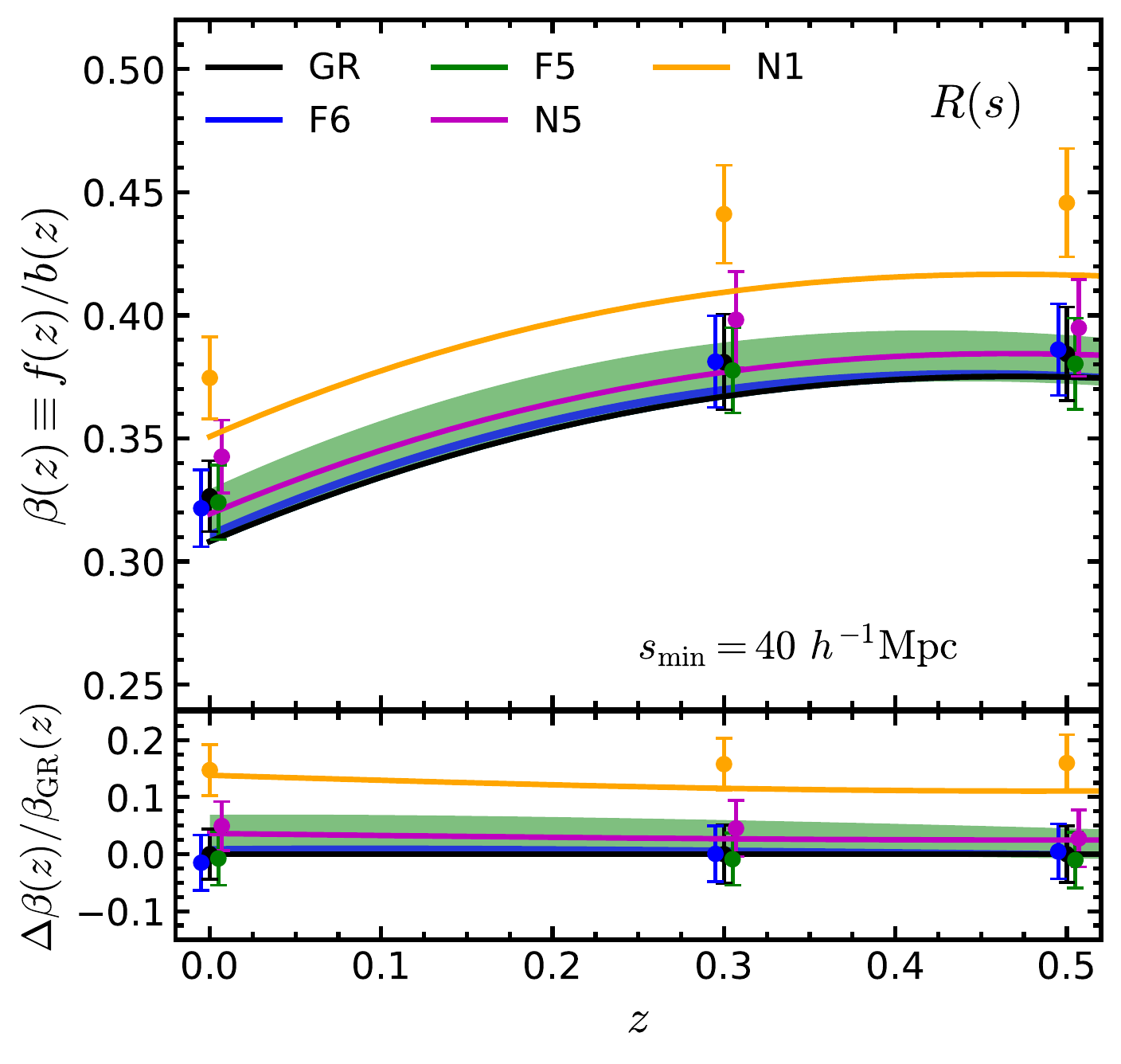}
	\caption{{\it Left panel}: Similar to the third row of Figure~\ref{fig:beta_lin_s1} but now we have used the full covariance matrix from the nonlinear Markov chains to estimate the best-fit values and error bars for $\beta$ by minimising the $\chi^2$ described in the text. {\it Middle panel}: the same as the left panel, but the estimator used in the constraint is $Q(s)$. {\it Right panel}: the same as the previous panels, but using the $R(s)$ estimator.}
	\label{fig:cov}
\end{figure*}

\subsection{The impact of covariance matrix}
\label{append:cov_mat}

In the constraints based on linear perturbation theory in Section \ref{sec:Kaiser}, the error bars in the $\chi^2$ minimisation were obtained as the standard deviations from the 5 realisations with 3 different LOS  of redshift-space galaxy catalogues for each model, which neglects the correlation between the different $s$ bins. Here we would like to check the impact of including such correlations on the parameter inference for $\beta$.

To this end we have redone the fitting of $\beta$ by using the covariance matrix from the Gaussian approximation instead of the standard deviations from the 15 realisations in Eq.~\eqref{eq:chi2}. We consider two estimators predicted using the linear Kaiser model, clustering wedges and $Q(s)$ (see Sec.~\ref{sec:Kaiser} for details), for this test. Since we are taken the covariance matrix, we minimise a $\chi^2$ given by
\begin{equation}
\chi^2(\beta) = [E(s_i) - E^{\rm th}(s_i;\beta)]^{T} C^{-1(E)}_{ij} [E(s_j) - E^{\rm th}(s_j;\beta)]\,,
\end{equation}
where $E(s)$ is the measured estimator, $E^{\rm th}(s;\beta)$ is the theoretical prediction from the linear model and $C^{(E)}_{ij}$ is the covariance matrix for each estimator.

For linear clustering wedges, we use the covariance matrix from their non-linear analogue. In the left panel of Fig.~\ref{fig:cov} we can find the results of this exercise for scales $s_{\rm min} = 40\Mpch$. The result shows substantial difference from that displayed in the lower right panel of Fig.~\ref{fig:beta_lin_s1}.
This suggests that the exact way to estimate the error budget can have a non-negligible impact on the calculation of $\chi^2$, and consequently on the estimation of $\beta$. As we found in Sec.~\ref{sec:Kaiser}, the best-fitting $\beta$ values when $s_{\rm min} = 40\Mpch$ are closer to the true values than the results at $s_{\rm min} = 20\Mpch$. 
However, these estimations are once again well below to the true values of $\beta$ as measured from the mock galaxy catalogues.
One of the explanation for this behaviour could be due to the correlation between the error and the data itself. As pointed out by \citep{Dodelson:2013cov} when the error is estimated from data and especially when it is positively correlated with the data, the inferred parameters are likely to be biased high. Also we have to consider the fact that the 15 realizations have certain overlap among each other. Further test should be done in the future. Again, the relative differences between MG and GR are consistent with the previous findings in linear and nonlinear models.

We tried the same test for $Q(s)$, by using the Kaiser formula to predict its theoretical value and the same covariance matrix used for the nonlinear model to estimate the corresponding $Q(s)-$covariance matrix. Starting from the Gaussian covariance matrix for the correlation multipoles, we applied a basis transformation and obtained a covariance for the $Q(s)$ estimator, 
\begin{equation}\label{eqn:cov_basis_transformation}
C_{ij}^{Q} \equiv \sum_{l,m}\frac{\partial Q_i}{\partial V_{l}} \frac{\partial Q_j}{\partial V_{m}} C_{lm},
\end{equation}
where $C_{lm}$ is the covariance matrix for $[\xi_0, \xi_2]$ and ${\bf V} = [\xi_{0}, \xi_{2}]$ is the data vector. These tests showed similar qualitative behaviour to the case of using correlation function wedges discussed above, with more biased constraints on $\beta$, and the results are shown in the middle panel of Figure~\ref{fig:cov}.

Following a similar procedure to the $Q(s)$ estimator, we generated a covariance matrix for $R(s)$.
The covariance matrix for the $R(s)$ estimator requires the knowledge of the covariance between the real space correlation function and the redshift space monopole. To achieve that, first, we modified the Eq.~(15) in \citet{Grieb:2016cov} by inserting the product of power spectrum both in real and redshift space. Second, a basis transformation similar to Eq.~\eqref{eqn:cov_basis_transformation} is applied, 
\begin{equation}
C_{ij}^{R} \equiv \sum_{p,q}\frac{\partial R_i}{\partial U_{p}} \frac{\partial R_j}{\partial U_{q}} C_{pq}\,,
\end{equation}
where $C_{pq}$ is the covariance matrix for $[\xi_0, \xi_r]$ with data vector ${\bf U} = [\xi_{0}, \xi_{r}]$.
The biggest impact of the covariance matrix is the reduction of the error bar for all models at all redshifts (see right panel of Fig.~\ref{fig:cov}). The best-fitting $\beta$ values are higher than the fiducial values because there is an offset between the simulation measurements and the theoretical expectations (see left panel of Fig.~\ref{fig:RQ_sim}).

On the other hand, we have checked explicitly (not shown here) including correlations between different $s$ bins, i.e., the non-diagonal elements of the covariance matrix, leads to small changes in the best-fitting $\beta$ values.

\subsection{Posterior distributions of parameters}
\label{append:nuisance_params}

In the discussions in the main text, we have mainly focused on the constraints and posterior distribution of the parameter $\beta$. However, constraints on the other, nuisance, parameters could also be of interest, not only because they can help us to understand/interpret the results, but also because some of these parameters are physically meaningful quantities which may be affected by modified gravity. 

Figure~\ref{fig:2dcontour} shows the posterior distribution of the parameters in the MCMC fit. The different colours correspond to different cosmological models (following the same colour scheme as used in all the other plots). In these MCMC runs, all parameters including the AP parameters $q_\perp,q_\parallel$ were allowed to vary freely. The estimators used are the three multipole moments $\xi_l$, with $s_{\rm min}=40\Mpch$. All results are at $z=0.5$. We can see that $f\sigma_8$ shows by far the largest difference between the different gravity models, while most other parameters are fairly similar in all models. We have also checked the same figure from using the three wedges ($\xi_w$) as the estimators, and found the resulting posterior distributions of all parameters to be nearly identical (not shown here).

For the case of $s_{\rm min}=20h^{-1}$Mpc, we found that using correlation function wedges ($\xi_w(s)$) and multipole moments ($\xi_l(s)$) can lead to quite distinct posterior distributions for some parameters, in particular $b_2$ and $a_{\rm vir}$ (see Fig. \ref{fig:2dcontoursmin20}). This is not surprising given that the two estimators differ by the small-scale information they contain, which are most relevant to these two parameters. The constraints on the other parameters are more or less consistent between the two estimators. Comparing Figs.~\ref{fig:2dcontour} and \ref{fig:2dcontoursmin20}, we can see that (i) the constraint $f\sigma_8$ is higher in the case of $s_{\rm min}=20\Mpch$, similar to what was found in Figure \ref{fig:beta_freeAP} and by \citet{Barreira:2016ovx}, and (ii) the uncertainties in the parameter constraints are smaller in the case of $s_{\rm min}=40\Mpch$, reflecting the fact that more information (on smaller scales) is used. These indicate the importance of using a more accurate model for the theoretical predictions for parameter constraints and inferences.

We have also tested the effects of fixing the AP parameters in the MCMC fitting, and found its effect on the nuisance parameters is much smaller than that of using different estimators (multiples vs. wedges). Regarding the $\beta$ parameters the difference of free/fixing AP parameter is within one $\sigma$ for different models over the redshifts. The results are not shown here for simplicity.

\begin{figure*}
\centering
\includegraphics[width=1.0\textwidth]{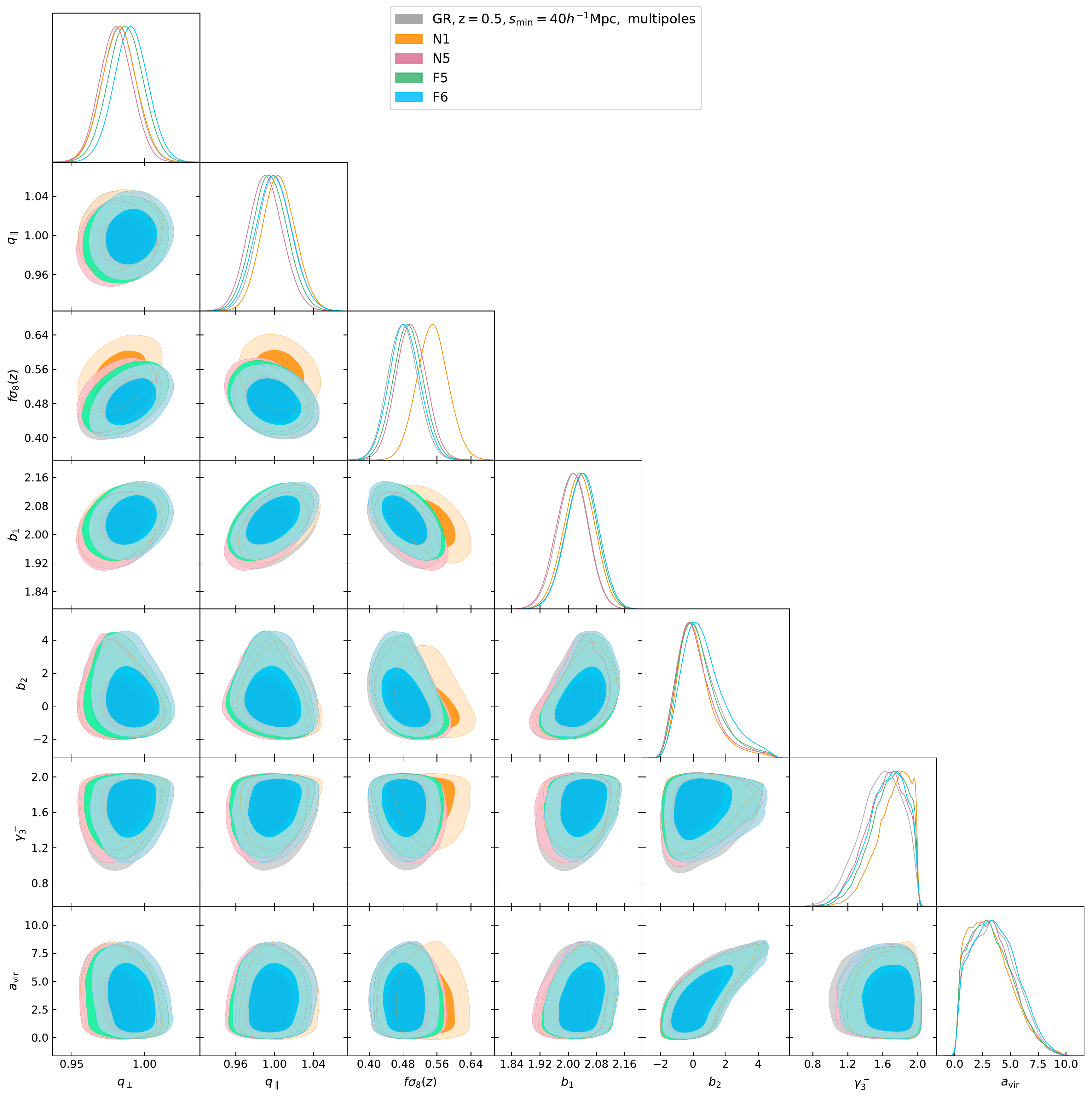}
	\caption{Posterior distribution of the parameters using three multipoles $\xi_{\ell= 0, 2, 4}$ with a minimum fitting range $s_{\rm min} = 40 \Mpch$ for different cosmological models. The distribution is evaluated at redshift $z=0.5$. The contours represent the $68\%$ (darker region) and $95\%$ (lighter region) confidence level.}
	\label{fig:2dcontour}
\end{figure*}

\begin{figure*}
\centering
\includegraphics[width=1.0\textwidth]{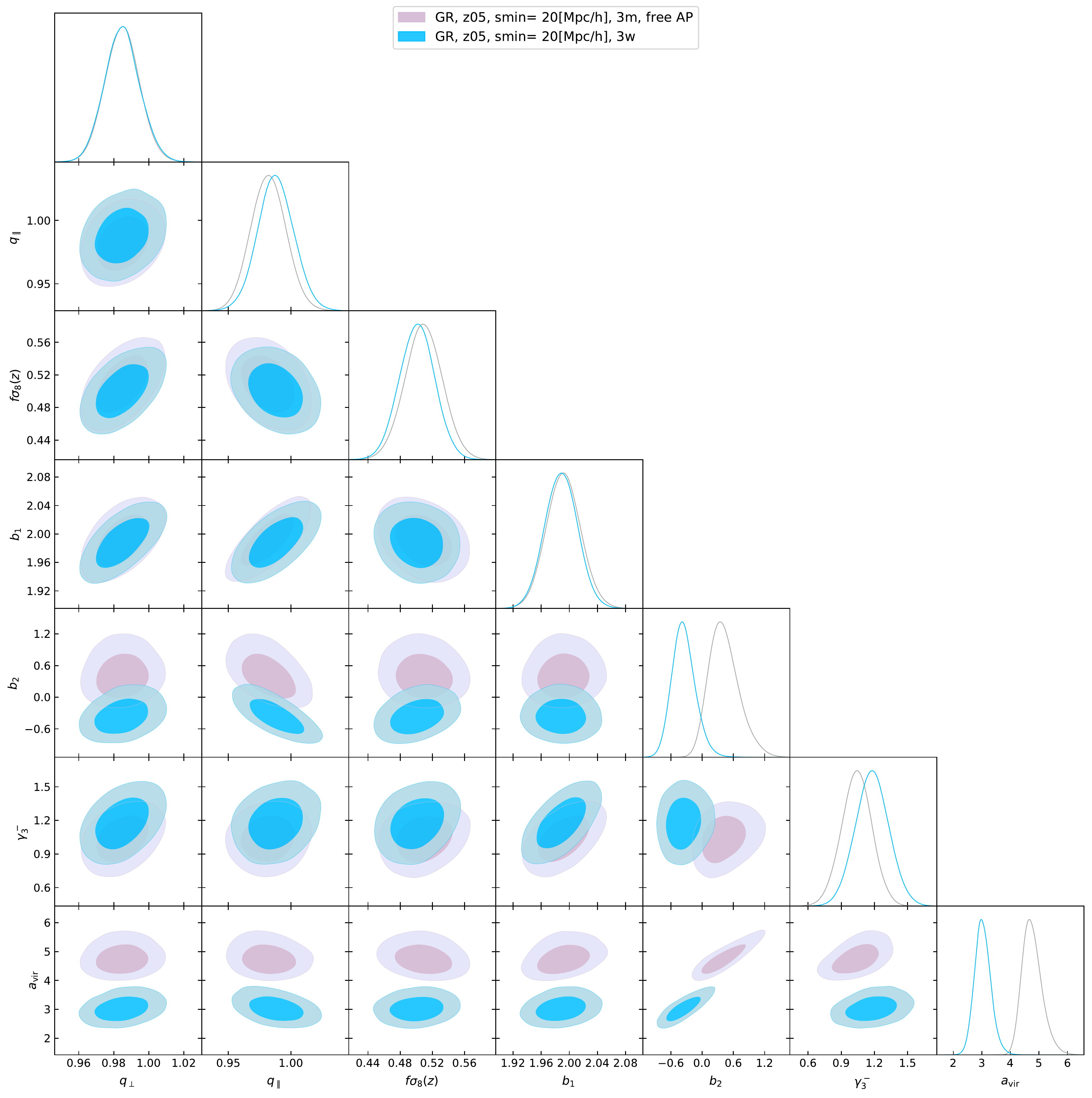}
	\caption{Posterior distribution of parameters using three multipole moments ($\xi_l(s)$, purple colour) and three correlation function wedges ($\xi_w(s)$, blue colour). The dark and light shaded regions are respectively the 1$\sigma$ and 2$\sigma$ contours, and the 1D marginalised distributions for the different parameters are shown as curves. The results are from the MCMC chains for GR at $z=0.5$, where the AP parameters were left free to vary, and the minimum scale used for the fitting was $s_{\rm min}=20h^{-1}$Mpc.}
	\label{fig:2dcontoursmin20}
\end{figure*}


\bsp	
\label{lastpage}
\end{document}